\documentclass[acmsmall,nonacm]{acmart}
\usepackage{amsmath,amsfonts}
\usepackage[ruled,vlined]{algorithm2e}
\usepackage{adjustbox}
\usepackage{graphicx}
\usepackage{textcomp}
\usepackage{xcolor}
\usepackage{physics}
\usepackage{wrapfig}
\usepackage{subcaption}
\usepackage{diagbox}
\usepackage{multirow}
\usepackage{multicol}
\usepackage{makecell}
\usepackage{enumitem}
\usepackage{hyperref}
\usepackage{bm}
\usepackage[normalem]{ulem}
\usepackage{stmaryrd}   
\hypersetup{}

\newif\iffinal
\finalfalse
\iffinal
  
  \newcommand{\deleted}[1]{}
\else
  
  \newcommand{\deleted}[1]{\textcolor{red}{\sout{#1}}}
\fi

\DeclareMathOperator*{\argmin}{arg\,min}

\newtheorem{hypothesis}{Hypothesis}[section]
\newtheorem{theorem}{Theorem}[section]

\newtheorem{lemma}{Lemma}[section]
\newtheorem{remark}{Remark}[section]
\newtheorem{definition}{Definition}[section]
\newtheorem{proposition}{Proposition}[section]

\newcommand{\myCompilerName}{FusionQ} 
\newcommand{\myCompilerNameSpace}{FusionQ }

\begin{document}
\raggedbottom
\title{\myCompilerName: A Generic and Optimal Decoding Framework for Ultra-Low-Latency Quantum Error Correction}

\author{Junyu Zhou}
\affiliation{%
  \institution{Microsoft Quantum}
  \city{Redmond}
  \country{USA}
}
\affiliation{%
  \institution{University of Pennsylvania}
  \city{Philadelphia}
  \country{USA}}
\email{junyuzh@engineering.upenn.edu}

\author{Sean Garner}
\affiliation{%
  \institution{Microsoft Quantum}
  \city{Redmond}
  \country{USA}
}
\affiliation{
  \institution{University of Washington}
  \city{Seattle}
  \country{USA}
}
\email{seangarn@uw.edu}

\author{Aarthi Sundaram}
\affiliation{%
  \institution{Microsoft Quantum}
  \city{Redmond}
  \country{USA}
}
\email{Aarthi.Sundaram@microsoft.com}

\author{Gushu Li}
\affiliation{%
  \institution{University of Pennsylvania}
  \city{Philadelphia}
  \country{USA}
}
\email{gushuli@engineering.upenn.edu}

\author{Yue Wu}
\affiliation{%
  \institution{Microsoft Quantum}
  \city{Redmond}
  \country{USA}
}
\email{yuewu4@microsoft.com}

\begin{abstract}
Quantum error correction (QEC) decoding has traditionally been organized around
fixed decoding instances: a monolithic space-time detector error model (DEM), or
a predetermined temporal window, is handed to a backend decoder after enough
syndrome data has arrived.  This paper presents \myCompilerName, a fusion
decoding framework that changes this paradigm by deriving local decoding tasks
from the observed syndrome and then fusing their results into a single global
correction.  Instead of forcing every shot through the same global or windowed
problem, \myCompilerNameSpace builds open-boundary local DEMs around fired
detector clusters and expands the corresponding footprints only when a local
optimum uses virtual boundary errors. 
\myCompilerNameSpace is generic in the decoder backend: the scheduler requires
only a local solution and a boundary-use flag, so exact solvers, matching-style
decoders on graphlike DEMs, and heuristic decoders can share the same runtime.
We prove a compositional optimality theorem: with an exact minimum-weight local
backend and boundary-free publication, fusing the published local decoding
results yields a global most-likely error (MLE) solution of the original DEM. Our evaluation shows that, with an exact backend, \myCompilerNameSpace matches
the monolithic MLE result on every shot.  Across decoder backends,
\myCompilerNameSpace substantially reduces decoding latency by
overlapping local decoding with hardware syndrome acquisition.
\end{abstract}

\maketitle 

\section{Introduction}
\label{sec:intro}

Quantum error correction (QEC)~\cite{gottesman1997stabilizer,dennis2002topological, shor1995scheme} is a foundational requirement for scalable
fault-tolerant quantum computation~\cite{chow2014implementing,aasen2025roadmap,bluvstein2026fault,paetznick2026improved}.
By encoding logical qubits into redundant physical degrees of freedom and
repeatedly measuring stabilizer checks, QEC systems produce a stream of syndrome
information that reveals the effects of physical faults.  A decoder must infer
an appropriate correction, or Pauli-frame update, from this stream.  Consequently,
decoding is not merely an offline post-processing task: it is an online runtime
component whose latency can directly constrain the execution rate of a
fault-tolerant quantum program~\cite{terhal2015quantum,google2025quantum}.

The detector error model (DEM)~\cite{gidney2021stim} is the standard abstraction
we use to formulate the decoding problem.  A DEM represents a noisy QEC
circuit as a weighted hypergraph in which vertices are detectors and hyperedges
are elementary fault mechanisms.  Given an observed syndrome, most-likely-error
(MLE) decoding asks for a minimum-weight set of fault mechanisms that produces
it.  Different backend decoders solve or approximate this objective under
different structural assumptions: for example, matching decoders such as MWPM target
graphlike DEMs~\cite{higgott2022pymatching}, search-based solvers such as Tesseract support general DEM instances~\cite{beni2025tesseract}, and heuristic decoders such as BP-OSD and
relay-BP approximate the MLE objective for improved scalability~\cite{roffe2020decoding,muller2025improved}.
This makes the DEM a natural interface for a decoding framework that should be
generic in its backend solver.

The central systems challenge is latency.  Existing decoding architectures are
largely organized around fixed problem instances.  A monolithic decoding strategy preserves
the full global MLE objective, but it must process a large space-time DEM and typically cannot start the decoding process until the entire syndrome volume has been accumulated.  Parallel window decoding mitigates this bottleneck by partitioning the DEM along the time dimension~\cite{skoric2023parallel,tan2023scalable,bombin2023modular}.
Lattice surgery~\cite{horsman2012surface, fowler2018low, zhou2026topols} and code surgery~\cite{williamson2026low, he2025extractors, zhou2026genecs} have emerged as central strategies for implementing fault-tolerant logical operations across a broad range of quantum error-correcting codes. Specialized spatial decompositions have also been developed for surface-code lattice surgery~\cite{lin2025spatially}. However, these decompositions rely on the particular geometric structure of the surface code and do not directly provide a decomposition for code surgery DEMs of general stabilizer codes~\cite{williamson2026low,zhou2026genecs}, whose detector hypergraphs may lack a regular spatial structure. Consequently, in the general heterogeneous QEC architecture with code surgery, a temporal window may still span a large decoding instance and leave substantial
post-acquisition latency.

\begin{figure}[t]
    \centering
    \includegraphics[width=0.73\linewidth]{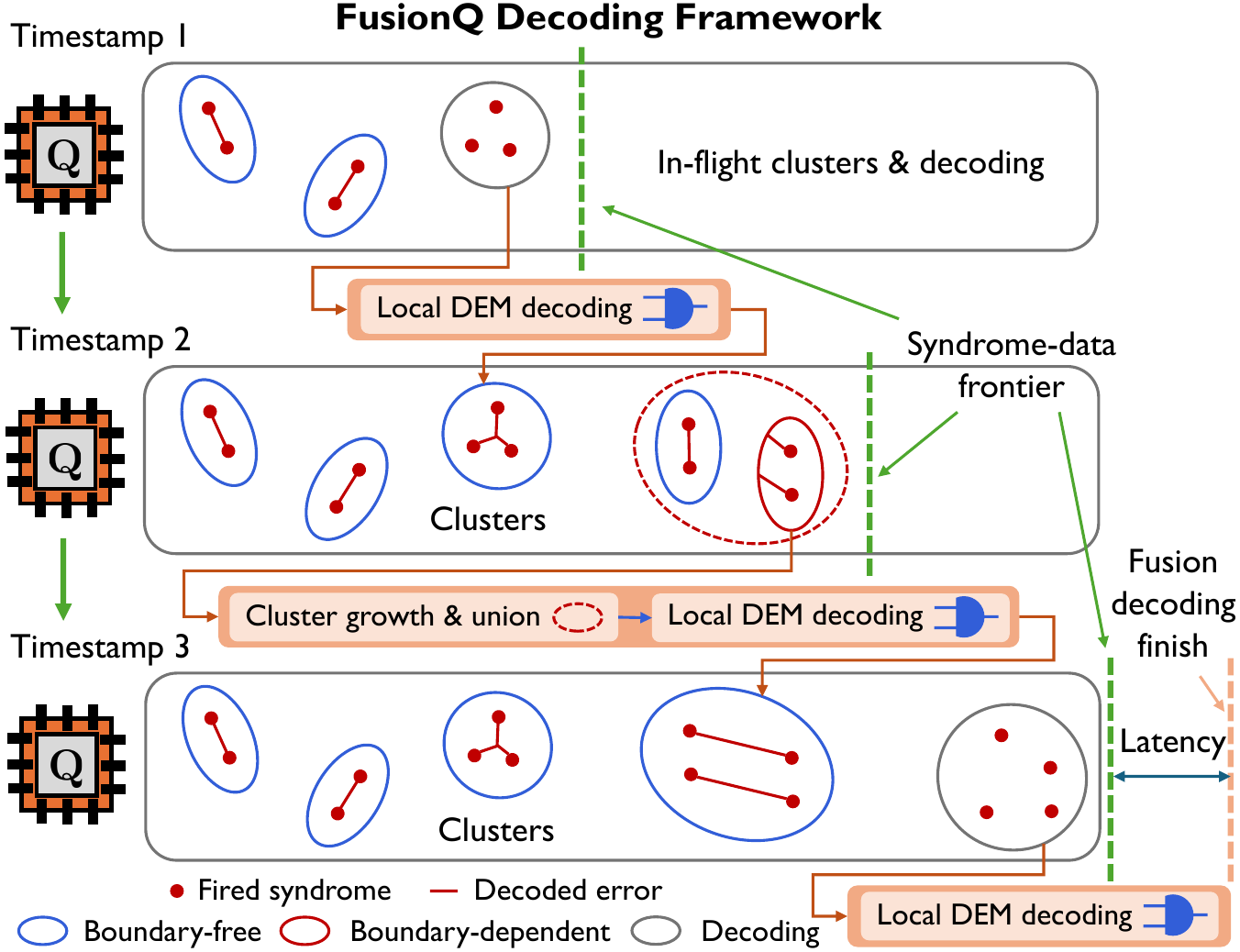}
    \caption{Online fusion decoding across three timestamps. Red dots are fired detectors, and solid red segments are decoded errors. Blue outlines mark boundary-free clusters, red outlines mark boundary-dependent clusters, and gray outlines mark clusters being decoded. The green dashed lines show the frontier of available syndrome data. At timestamp 2, a boundary-dependent result triggers footprint growth (dashed red outline), union with a neighboring cluster, and another local DEM solve. The merged cluster is boundary-free at timestamp 3. Latency is measured from the final syndrome-data frontier to decoding completion (orange dashed line).}
    \label{fig:overall}
\end{figure}

This paper is motivated by a different organizing principle:
\textbf{decoding regions should be derived dynamically from the observed
syndrome}.  At the low physical error rates (below $10^{-3}$) targeted by near-term fault-tolerant
architectures~\cite{loschnauer2025scalable, marxer2026above, bluvstein2026fault}, fired detectors occupy only a small fraction of the full
space-time volume and often form localized components.  This sparsity suggests a
paradigm shift: instead of defining the decoding decomposition before execution,
the runtime system should derive local decoding tasks from the fired syndrome
itself, expand only the tasks whose local evidence is insufficient, and combine
their decoding results only when the corresponding local problems can be
certified independently.

We present \myCompilerName, a dynamic fusion framework for low-latency QEC
decoding shown in
Figure~\ref{fig:overall}.  The overview follows one decoding task across
multiple ordered timestamps: local DEM decoding
begins as soon as the relevant syndrome data are available, boundary-dependent
solutions trigger footprint growth and union, and publication waits until the live
local solutions are boundary-free.  The key idea is to turn decoding from a fixed decomposition of the
full space-time DEM into a runtime decomposition induced by the observed
syndrome.  Rather than invoking a backend decoder on one monolithic or
predetermined windowed instance, \myCompilerNameSpace constructs local decoding
tasks around the detector regions that are actually activated, schedules these
tasks as soon as their detector outcomes are available, and fuses their
boundary-free decoding results into a single global correction.

\textbf{First}, \myCompilerNameSpace localizes the observed syndrome by grouping
fired detectors into one-hop connected fired cores and initializing a
radius-one cluster footprint around each core.  For each footprint,
\myCompilerNameSpace constructs an open-boundary local DEM whose boundary
detectors are equipped with zero-weight virtual boundary errors.  These virtual
errors allow the local optimum to reveal whether the current footprint is large
enough to explain the syndrome assigned to its interior.

\textbf{Second}, \myCompilerNameSpace certifies local answers through a
boundary-free condition.  A boundary-free optimum gives a tentative local
decoding result, but that result is not published in isolation: a later
boundary-dependent task may grow into its footprint and require re-decoding on
the combined local problem.  \myCompilerNameSpace publishes only at an all-clear
fixed point, where every live footprint has a boundary-free local solution and
every pair of live footprints intersects only on boundary detectors.

\textbf{Third}, \myCompilerNameSpace resolves boundary-dependent tasks by growing
their anchored footprints and taking detector-set unions with any live footprint
whose overlap is not purely boundary-boundary.  After each footprint union,
\myCompilerNameSpace recomputes the boundary, interior, and local DEM from the
original global DEM, then invokes the backend again on the recomputed local
problem.  This grow/union/re-decode loop continues until the re-decoded solution
is boundary-free.

Our experimental results show that \myCompilerNameSpace achieves ultra-low
post-acquisition latency across diverse code families, decoder backends, and QEC
workloads, with reductions reaching orders of magnitude over monolithic
and parallel window decoding. With syndrome data produced at a rate of one QEC
round every $5\,\mathrm{ms}$, its mean post-acquisition latency remains below
$5\,\mathrm{ms}$ in every representative configuration; in many configurations,
even the p95 or p99 latency---the latency within which $95\%$ or $99\%$ of shots
complete decoding---remains below $5\,\mathrm{ms}$. We further scale the number of code
blocks participating in code surgery, the code distance, and the QEC circuit
depth. Across these growing spatial and temporal workloads,
\myCompilerName's post-acquisition critical path grows far more slowly than
monolithic decoding and remains nearly depth-independent at fixed distance.
Together, these results show that \myCompilerNameSpace maintains the same or better
correction quality while turning decoding from an end-of-acquisition bottleneck
into a streaming runtime that can keep pace with syndrome production.

The major contributions of this paper are as follows.
\begin{enumerate}[leftmargin=*]
\item \textbf{A generic fusion-decoding abstraction.}
We introduce a solver-parametric framework that constructs local decoding tasks
from the observed syndrome and fuses their results, allowing any backend that
solves or approximates the MLE objective to share the same runtime.

\item \textbf{A global optimality guarantee.}
We define open-boundary local DEMs and prove a compositional optimality theorem: if the local backend is exact, the published local solutions are boundary-free, and the published footprints satisfy the boundary-only intersection condition, then fusing those local solutions yields a global minimum-weight solution of the original DEM.

\item \textbf{An ultra-low-latency decoding architecture.}
We develop an online execution model that separates local DEM construction, backend decoding, footprint growth and union, and
global publication, providing a practical architecture for real-time QEC
decoding.
\end{enumerate}

\section{Preliminary}
\label{sec:prelim}

This section reviews the background needed for the paper: quantum error correction
codes, detector error model, and most-likely error decoding. For a broader introduction
to quantum computing, we recommend~\cite{nielsen2001quantum}.

\subsection{Quantum Error Correction Codes}
\label{subsec:qec}

\begin{figure}[t]
    \centering
    \includegraphics[width=0.95\linewidth]{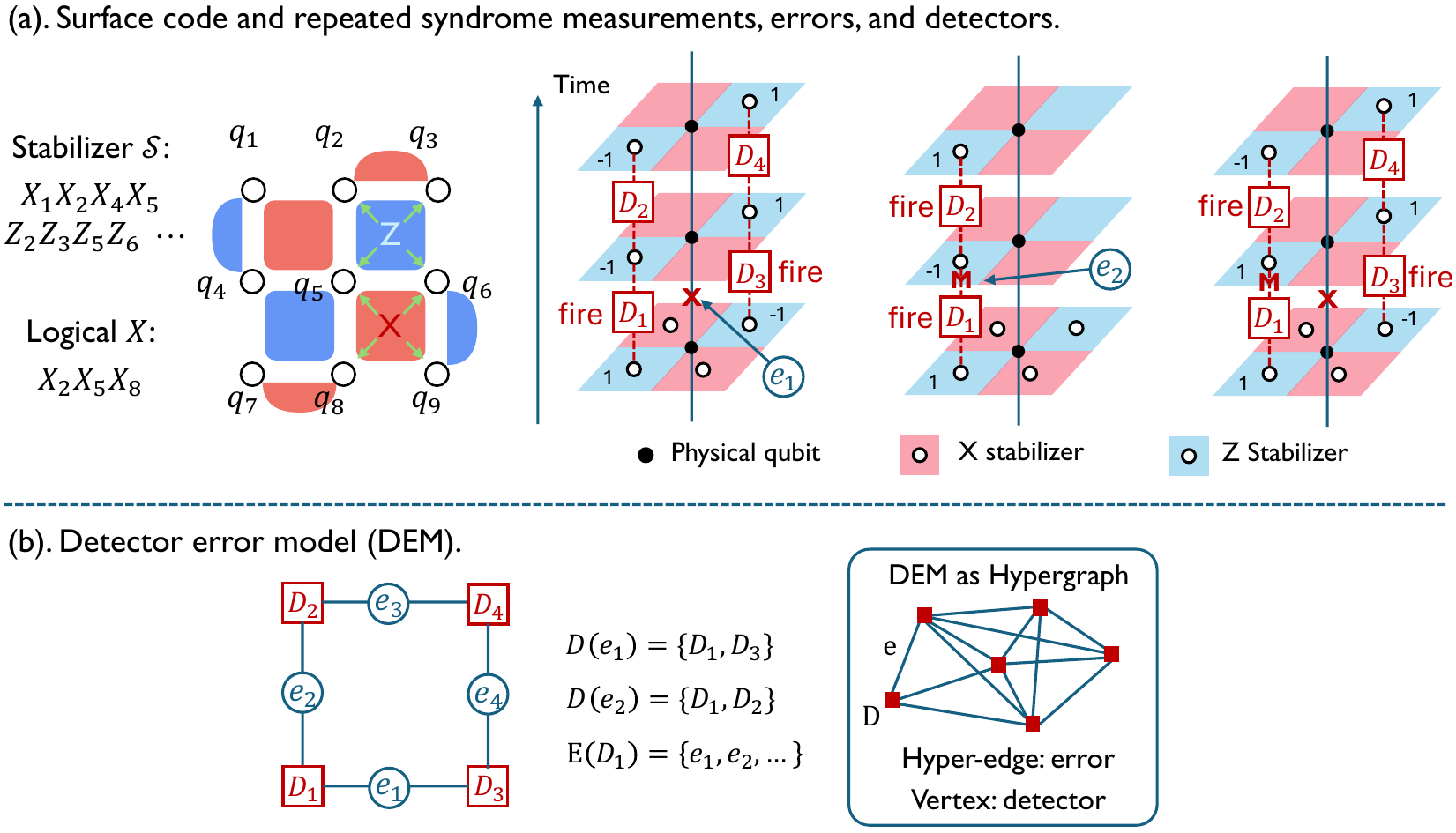}
    \caption{(a) A $\llbracket 9,1,3 \rrbracket$ surface code and repeated
    syndrome measurements, with time increasing upward. A data error $e_1$
    fires $D_1,D_3$; a measurement error $e_2$ fires $D_1,D_2$; together,
    only $D_2,D_3$ fire. (b) Detectors are vertices and error mechanisms are
    hyperedges. The four-detector example is graphlike: each shown error touches
    two detectors, while a general DEM error may touch more than two.}
    \label{fig:surface-code}
    \label{fig:dem}
\end{figure}

Quantum error correction (QEC)~\cite{gottesman1997stabilizer,dennis2002topological} protects logical quantum information by encoding it
redundantly into a larger set of physical qubits. A quantum error correction code is
characterized by a set of parity constraints, commonly represented by \emph{stabilizer
checks}, that define the valid code space. In the absence of errors, these checks
produce deterministic outcomes. Physical errors may flip one or more check outcomes,
thereby providing information about where errors may have occurred without directly
measuring the encoded logical state.

Figure~\ref{fig:surface-code}(a) shows a distance-three surface code with nine data
qubits. Each colored plaquette defines a stabilizer check acting on the four adjacent
qubits. For example, the left red plaquette corresponds to the $X$-type stabilizer
$X_1X_2X_4X_5$, while the right blue plaquette corresponds to the $Z$-type stabilizer
$Z_2Z_3Z_5Z_6$. There are also weight-2 stabilizers at the boundary, such as
$Z_1Z_4$. A logical Pauli operator commutes with every stabilizer but is not itself a
stabilizer. As illustrated, $X_2X_5X_8$ is one representative logical $X$ operator.

\subsection{Detector Error Model}
\label{subsec:dem}

In fault-tolerant quantum computing~\cite{gottesman1998theory}, stabilizer measurements are repeatedly performed
over multiple rounds. Rather than interpreting each measurement outcome independently,
error correction considers correlations between measurement outcomes across space and
time. These correlations define \textbf{detectors}, whose outcomes are deterministic in
the absence of errors. A detector \textbf{fires} when an error causes the corresponding
deterministic relation to be violated.

Figure~\ref{fig:dem}(a) illustrates this process using repeated syndrome extraction in
the surface code. The same stabilizers are measured in consecutive rounds, and a
detector checks the parity between adjacent measurement outcomes. Consider the physical
$X$ error $e_1$. Since it anti-commutes with the two neighboring $Z$-stabilizer checks,
their outcomes flip across rounds, causing $D_1$ and $D_3$ to fire. In contrast, $e_2$
is a measurement error that flips a stabilizer outcome in only one round, producing two
consecutive changes and causing $D_1$ and $D_2$ to fire. When $e_1$ and $e_2$ occur
together, their effects combine modulo two: $D_1$ is flipped twice and therefore
cancels, leaving only $D_2$ and $D_3$ fired.

The \textbf{detector error model (DEM)}~\cite{gidney2021stim} abstracts these
error-detector relationships from the underlying QEC circuit. As illustrated in
Figure~\ref{fig:dem}(b), we represent a DEM as a weighted hypergraph $G =
(\mathcal{D},\mathcal{E})$, where the vertex set $\mathcal{D}$ contains detectors and
the hyperedge set $\mathcal{E}$ contains elementary error mechanisms. Each error
hyperedge connects exactly the detectors that it causes to fire. The representation is
generally a hypergraph because a single error mechanism may trigger
more than two detectors.

We use $D(e_i)$ to denote the set of detectors associated with an error mechanism $e$.
For example, Figure~\ref{fig:dem}(b) gives
$D(e_1) = \{ D_1, D_3 \}$, $D(e_2) = \{ D_1, D_2 \}$.
Conversely, $E(D_j)$ denotes the set of error mechanisms incident on detector $D_j$; for
example, $E(D_1) = \{ e_1, e_2, \ldots \}$.
Thus, $D(\cdot)$ maps an error hyperedge to its incident detector vertices, while
$E(\cdot)$ maps a detector vertex to its incident error hyperedges.

More generally, let $\mathcal{F} \subseteq \mathcal{E}$ denote an \textbf{error set}, i.e., a collection of error mechanisms that occur together. We extend $D(\cdot)$ from individual errors to error sets by defining:
\begin{equation*}
    D(\mathcal{F}) = \bigoplus_{e \in \mathcal{F}} D(e),
\end{equation*}
where $\oplus$ denotes the symmetric difference. Therefore, a detector belongs to
$D(\mathcal{F})$ if and only if it is triggered by an odd number of errors in
$\mathcal{F}$. For example, for $\mathcal{F} = \{ e_1, e_2 \}$,
\begin{equation*}
D(\mathcal{F}) = D(e_1) \oplus D(e_2)
= \{ D_1, D_3 \} \oplus \{ D_1, D_2 \} = \{ D_2, D_3 \},
\end{equation*}
which corresponds exactly to the rightmost example in Figure~\ref{fig:dem}(a), where the
two contributions to $D_1$ cancel.

\begin{wrapfigure}{R}{0.45\textwidth}
    \centering
    \includegraphics[width=0.45\textwidth]{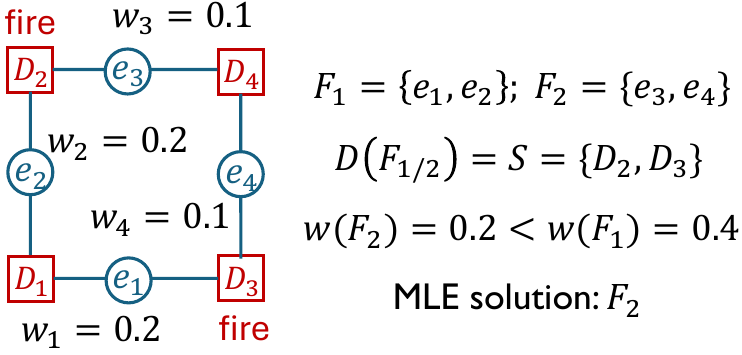}
    \caption{Most-Likely Error (MLE) decoding.}
    \label{fig:mle}
\end{wrapfigure}

Once a QEC circuit is represented by its DEM, decoding can be formulated entirely on
this hypergraph: given an observed set of fired detectors $\mathcal{S} \subseteq
\mathcal{D}$, the decoder seeks an error set $\mathcal{F} \subseteq \mathcal{E}$
satisfying $D(\mathcal{F}) = \mathcal{S}$. This paper studies the
most-likely-error objective over elementary mechanisms. The next section introduces how
the most likely such error set is identified.

\subsection{Most-Likely Error Decoding}
\label{subsec:mle}

Given an observed detector syndrome $\mathcal{S} \subseteq \mathcal{D}$, i.e., the set
of fired detectors, the decoding problem is to infer the error set that most likely
produced the observation. Recall that an error set $\mathcal{F} \subseteq \mathcal{E}$
produces the detector signature $D(\mathcal{F})$. We say that $\mathcal{F}$ is
\textbf{consistent} with the observed syndrome $\mathcal{S}$ if $D(\mathcal{F}) =
\mathcal{S}$. In \textbf{Most-Likely Error (MLE) decoding}~\cite{higgott2022pymatching,beni2025tesseract}, the decoder searches over
all consistent error sets and selects the most probable one.

Under an independent error model, where each elementary error mechanism $e_i$ occurs
independently with probability $p_i \in (0,1/2]$, the probability of an error set $\mathcal{F}$ is:
\begin{equation*}
    P(\mathcal{F}) = \prod_{e_i \in \mathcal{F}} p_i \prod_{e_i \notin \mathcal{F}}
(1 - p_i) \;\propto\; \prod_{e_i \in \mathcal{F}} \frac{p_i}{1 - p_i}.
\end{equation*}

This probabilistic optimization can be equivalently formulated as a
\textbf{minimum-weight decoding problem}. We assign each error mechanism $e_i$ the
log-likelihood weight:
\[
w_i = -\log \frac{p_i}{1 - p_i} ,
\]
and define the weight of an error set as:
\[
W(\mathcal{F}) = \sum_{e_i \in \mathcal{F}} w_i .
\]

Maximizing $P(\mathcal{F})$ is then equivalent to minimizing $W(\mathcal{F})$, so MLE
decoding can be written as
\begin{equation}
\label{eq:mle-argmin}
\mathcal{F}^{*} = \argmin_{\mathcal{F} \,:\, D(\mathcal{F}) = \mathcal{S}}
W(\mathcal{F}) .
\end{equation}

Figure~\ref{fig:mle} illustrates this optimization on a simple DEM. Suppose $D_2$ and
$D_3$ fire, so that $\mathcal{S} = \{ D_2, D_3 \}$. Two error sets are consistent with
this observation: $\mathcal{F}_1 = \{ e_1, e_2 \}$, whose shared detector $D_1$
cancels, and $\mathcal{F}_2 = \{ e_3, e_4 \}$, whose shared detector $D_4$ cancels,
giving $D(\mathcal{F}_1) = D(\mathcal{F}_2) = \mathcal{S}$. Their weights, however,
differ:
\[
W(\mathcal{F}_1) = w_1 + w_2 = 0.4 , \qquad
W(\mathcal{F}_2) = w_3 + w_4 = 0.2 ,
\]
so MLE decoding selects $\mathcal{F}_2$. This minimum-weight formulation serves as the
decoding objective throughout the remainder of this paper.

\section{Fusion Decoding Architecture}
\label{sec:architecture}

In this section, we present \myCompilerName, a fusion decoding architecture that solves
the global minimum-weight problem of Equation~\ref{eq:mle-argmin} as a collection of small
local problems, each decoded as soon as its syndrome data arrives.

\subsection{Motivation}
\label{subsec:motivation}

\begin{figure}[t]
    \centering
    \includegraphics[width=0.74\linewidth]{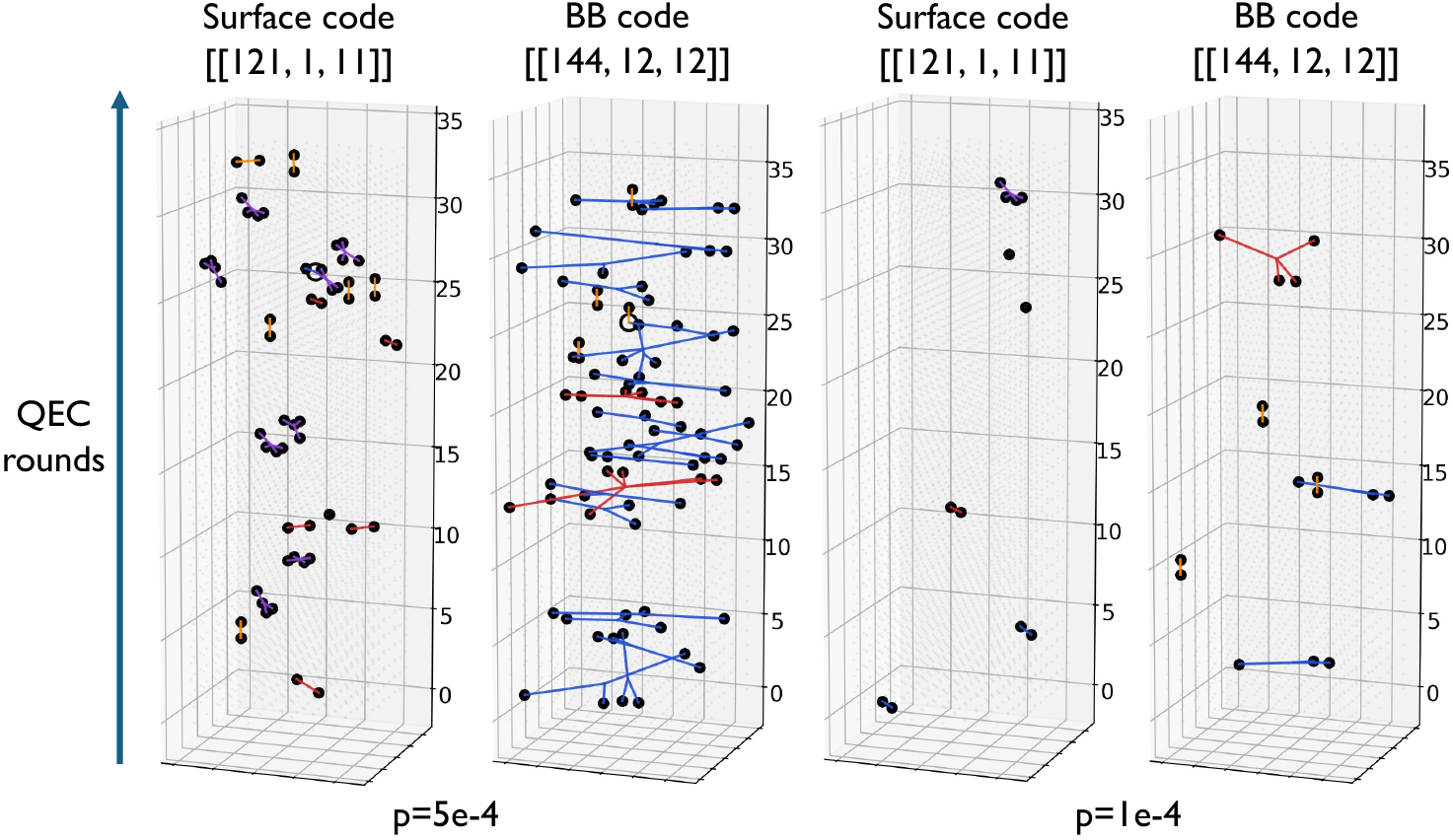}
    \caption{Typical detector syndrome at $p = 5\times10^{-4}$ and $p = 10^{-4}$.}
    \label{fig:motivation}
\end{figure}

At the physical error rates targeted by near-term fault tolerance ($p \lesssim
10^{-3}$), detection events are \emph{sparse} and \emph{separate in space--time}. The
overwhelming majority of measurement rounds are clean; faults appear as small, isolated
islands of activity in an otherwise quiet space--time volume, and distinct islands are typically
far apart in error-hop distance.


Figure~\ref{fig:motivation} illustrates this behavior for typical shots of the
surface code $\llbracket 121,1,11 \rrbracket$ and the bivariate-bicycle (BB) code $\llbracket 144,12,12 \rrbracket$ at $p = 5\times10^{-4}$ and $p = 10^{-4}$ respectively. Black dots denote fired detectors, plotted at their $(x,y,t)$ coordinates, and the faint background shows the full detector set. In both examples, the fired
detectors occupy only a small fraction of the space-time volume and form
localized islands rather than a single extended structure.

The figure also shows that, at the physical error rates used in these examples, most fired detectors are produced by isolated single-error mechanisms. There are a few local multi-error interactions: for example, in the
$p=5\times10^{-4}$ surface-code shot, two nearby error mechanisms may flip the
same detector, causing its parity to cancel so that it does not appear as a
fired detector. We mark such cancelled detectors with black circles. These
examples motivate treating the observed syndrome as a collection of mostly
localized components, while still allowing local regions to grow or fuse when
nearby error mechanisms interact.

A conventional \emph{monolithic} decoder nevertheless ingests the entire space--time
syndrome volume and solves one global optimization, spending most of its effort
propagating information across different regions. Even parallel window
decoding~\cite{skoric2023parallel}, which slices the volume along the time axis, still
processes the full spatial extent of the code over a window of $3d$ rounds, where $d$
is the code distance, so its smallest problem instance is fixed by the code size rather
than by the observed syndrome. This is the
inefficiency \myCompilerNameSpace removes: it localizes work to exactly the active
islands, and couples islands only when they are close enough to genuinely interact.

\subsection{Local DEM and Decoding}
\label{subsec:local}

\begin{figure}[t]
    \centering
    \includegraphics[width=\linewidth]{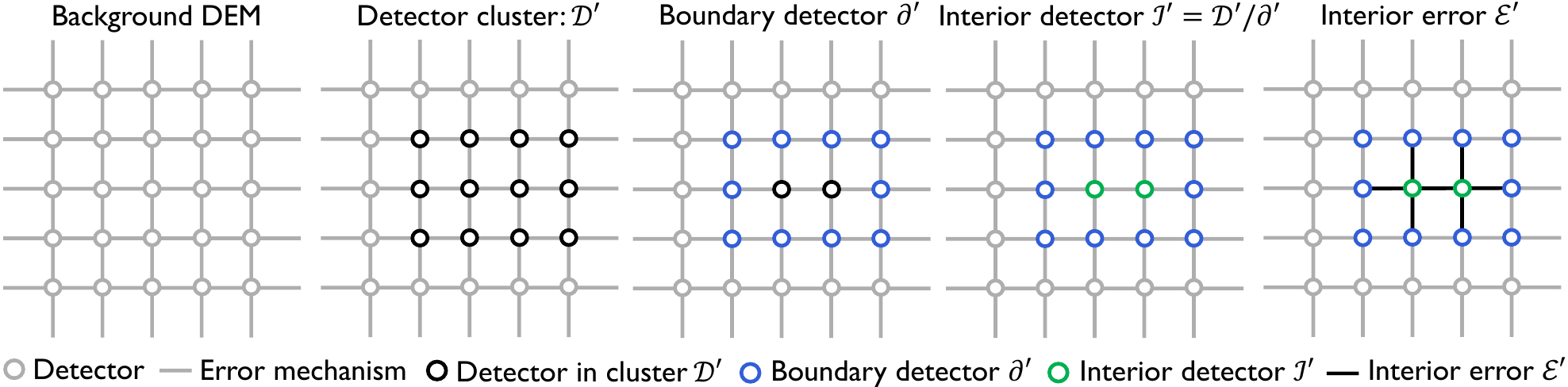}
    \caption{Detector cluster.}
    \label{fig:cluster}
\end{figure}

The unit of work in \myCompilerNameSpace is a \textbf{local DEM} built from a
\textbf{detector cluster}. Figure~\ref{fig:cluster} shows the process of constructing a
detector cluster, starting from the background DEM on the left, where grey circles are
detectors and grey lines are error mechanisms.

\begin{definition}[Detector cluster]
\label{def:cluster}
A \textbf{detector cluster} is a subset of detectors $\mathcal{D}' \subseteq
\mathcal{D}$. Its detectors split into the \textbf{boundary detectors}, those with at least one incident error mechanism escaping the cluster:
\begin{equation*}
\partial' = \{\, d \in \mathcal{D}' \mid \exists\, e \in E(d) \ \text{s.t.}\
D(e) \not\subseteq \mathcal{D}' \,\} ,
\end{equation*}
and the \textbf{interior detectors}, which is the complement of boundary detectors:
\begin{equation*}
\mathcal{I}' = \mathcal{D}' \setminus \partial'
= \{\, d \in \mathcal{D}' \mid \forall\, e \in E(d),\ D(e) \subseteq \mathcal{D}' \,\} .
\end{equation*}
The \textbf{interior errors} of the cluster are the error mechanisms incident to its interior
detectors,
\begin{equation*}
\mathcal{E}' = \bigcup_{d \in \mathcal{I}'} E(d).
\end{equation*}
\end{definition}

The four panels of Figure~\ref{fig:cluster} follow this construction in order: the
cluster $\mathcal{D}'$ in black, its boundary detectors $\partial'$ in blue, its
interior detectors $\mathcal{I}'$ in green, and its interior errors $\mathcal{E}'$ as
black lines. The two conditions in Definition~\ref{def:cluster} are exact negations, so
every detector of the cluster is either interior or boundary; and because an interior
detector has no escaping error mechanism, \textbf{every error mechanism in $\mathcal{E}'$ has its entire
detector footprint inside $\mathcal{D}'$}, so the cluster sees each of them in full. In addition, \textbf{the interior error $\mathcal{E}'$ is the only error set in the background DEM that adjacent to the interior detectors $\mathcal{I}'$}.

These definitions use only the incidence between detectors and error mechanisms, so although
Figure~\ref{fig:cluster} is drawn on a two-dimensional lattice, they apply unchanged to
a general DEM hypergraph, where a cluster boundary need not be a geometric surface.

A cluster on its own is not yet a decoding problem: an error mechanism reaching in from
outside can leave a defect on a boundary detector that no interior error can explain.
We therefore complete the cluster into a DEM of its own by attaching a free escape to
every boundary detector.

\begin{figure}[t]
    \centering
    \includegraphics[width=0.8\linewidth]{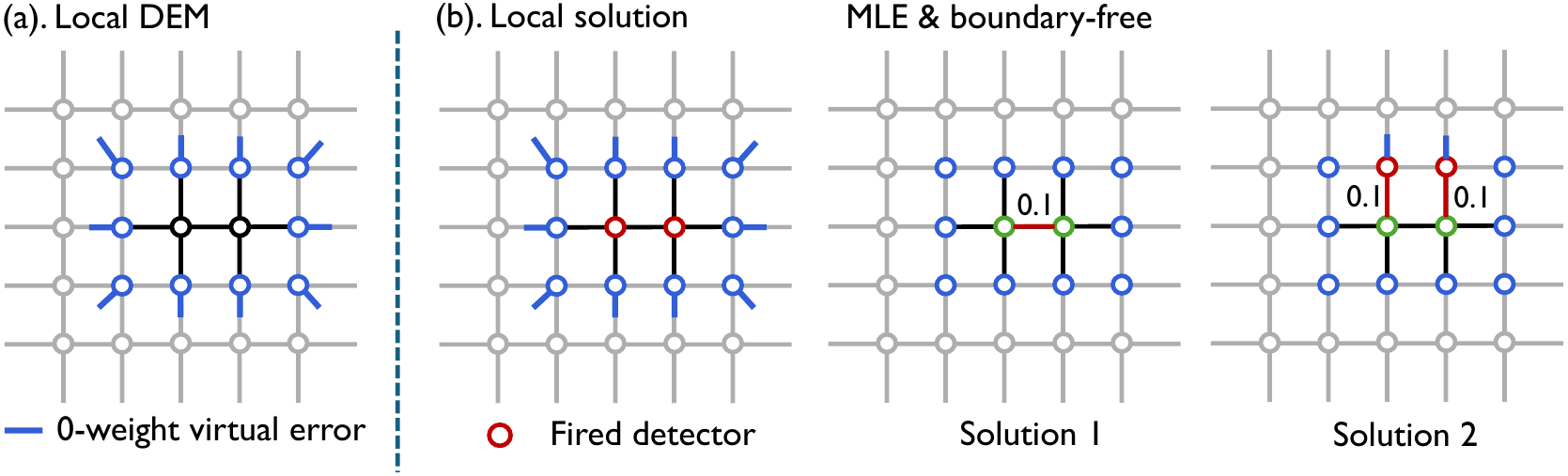}
    \caption{Local DEM and local decoding solution.}
    \label{fig:local-dem}
\end{figure}

\begin{definition}[Local DEM]
\label{def:local-dem}
The \textbf{local DEM} of a detector cluster $\mathcal{D}'$ is the DEM on the detector
set $\mathcal{D}'$ whose error mechanisms are the interior errors $\mathcal{E}'$
together with the \textbf{virtual boundary errors}
\begin{equation}
\label{eq:virtual}
\mathcal{V}' = \{\, e^{virt}_d \mid d \in \partial' \,\} ,
\qquad
D(e^{virt}_d) = \{ d \} ,
\qquad
w(e^{virt}_d) = 0 ,
\end{equation}
one for each boundary detector, each firing only that detector and carrying zero weight.
\end{definition}

Figure~\ref{fig:local-dem}(a) shows a local DEM: the interior errors in black, and one
zero-weight virtual error, drawn in blue, on every boundary detector. Decoding it is the
same minimum-weight problem as before, now posed inside the local DEM.

\begin{definition}[Local decoding solution]
\label{def:local-solution}
Let $\mathcal{S}' = \mathcal{S} \cap \mathcal{I}'$ be the fired detectors of the cluster. An error set $\mathcal{F}' \subseteq \mathcal{E}' \cup \mathcal{V}'$ is \textbf{feasible} if $D(\mathcal{F}') = \mathcal{S}'$, and the \textbf{local decoding solution} is the minimum-weight feasible set:
\begin{equation}
\label{eq:local-mle}
\mathcal{F}'^{*} = \argmin_{\mathcal{F}' \subseteq \mathcal{E}' \cup \mathcal{V}'
\,:\, D(\mathcal{F}') = \mathcal{S}'} W(\mathcal{F}') .
\end{equation}
A solution is \textbf{boundary-free} if it uses no virtual boundary error, i.e.\ if
$\mathcal{F}' \cap \mathcal{V}' = \emptyset$, otherwise it is called \textbf{boundary-dependent}.
\end{definition}

Figure~\ref{fig:local-dem}(b) shows the same two fired detectors explained in two ways:
Solution~1 uses the single interior error joining them, for weight $0.1$, and touches no
virtual error; Solution~2 instead routes both defects out to the boundary with two
interior errors of weight $0.1$ each, absorbing the resulting boundary defects with
virtual errors. Solution~1 is therefore both minimum-weight and boundary-free, while
Solution~2 is neither.

\subsection{Fusion Decoding Theory}
\label{subsec:theory}

\begin{figure}[t]
    \centering
    \includegraphics[width=0.55\linewidth]{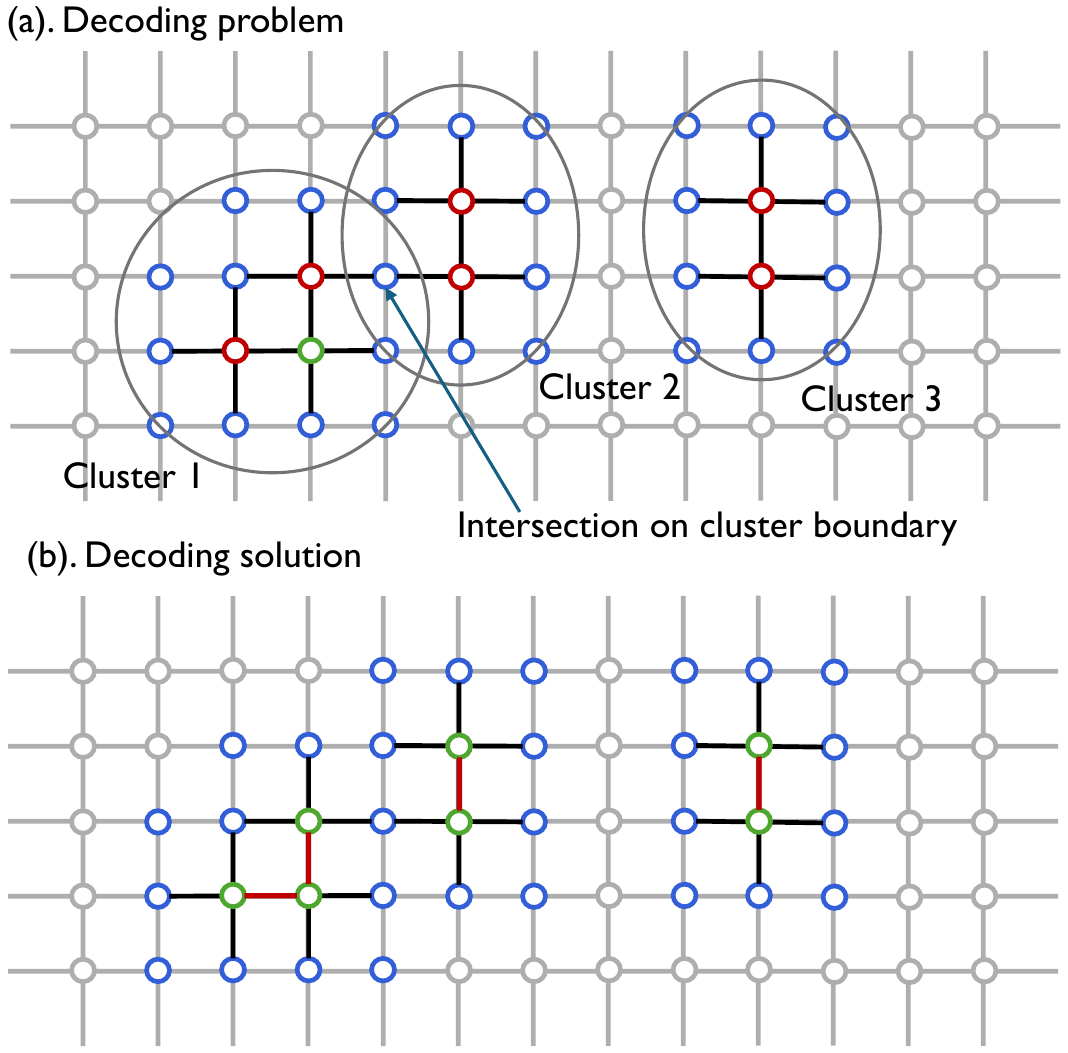}
    \caption{Fusion decoding.}
    \label{fig:fusion-theory}
\end{figure}

We now state the result that makes fusion decoding exact.

\begin{theorem}[Fusion decoding]
\label{thm:fusion}
Let $\mathcal{D}'_1, \ldots, \mathcal{D}'_k$ be detector clusters, with boundaries
$\partial'_i$, interiors $\mathcal{I}'_i$ and interior errors $\mathcal{E}'_i$, and
write $\mathcal{S}'_i = \mathcal{S} \cap \mathcal{I}'_i$ for the fired detectors each
one contains. Assume that $\mathcal{S}$ is feasible in the global DEM and that
all real error weights are non-negative. Suppose that
\begin{enumerate}[label=(\roman*),leftmargin=*]
\item the clusters interior cover the syndrome, $\mathcal{S} \subseteq \bigcup_i \mathcal{I}'_i$;
\item distinct clusters intersect only on boundary detectors,
      $\mathcal{D}'_i \cap \mathcal{D}'_j \subseteq \partial'_i \cap \partial'_j$
      for all $i \neq j$.
\end{enumerate}
Each cluster then has a local decoding solution $\mathcal{F}'^{*}_i$ in the sense of
Definition~\ref{def:local-solution}. If every $\mathcal{F}'^{*}_i$ is boundary-free,
their union is a global minimum-weight solution,
\begin{equation}
\label{eq:fusion}
\mathcal{F}^{*} = \bigcup_{i=1}^{k} \mathcal{F}'^{*}_i .
\end{equation}
\end{theorem}

Figure~\ref{fig:fusion-theory}(a) shows three clusters meeting these conditions: the
fired detectors sit in cluster interiors, and where two clusters overlap they share only
boundary detectors. The proof rests on the following observation, which is where
condition~(ii) does its work: clusters may share detectors, but they can never share an interior error mechanism.

\begin{figure}[t]
    \centering
    \includegraphics[width=0.55\linewidth]{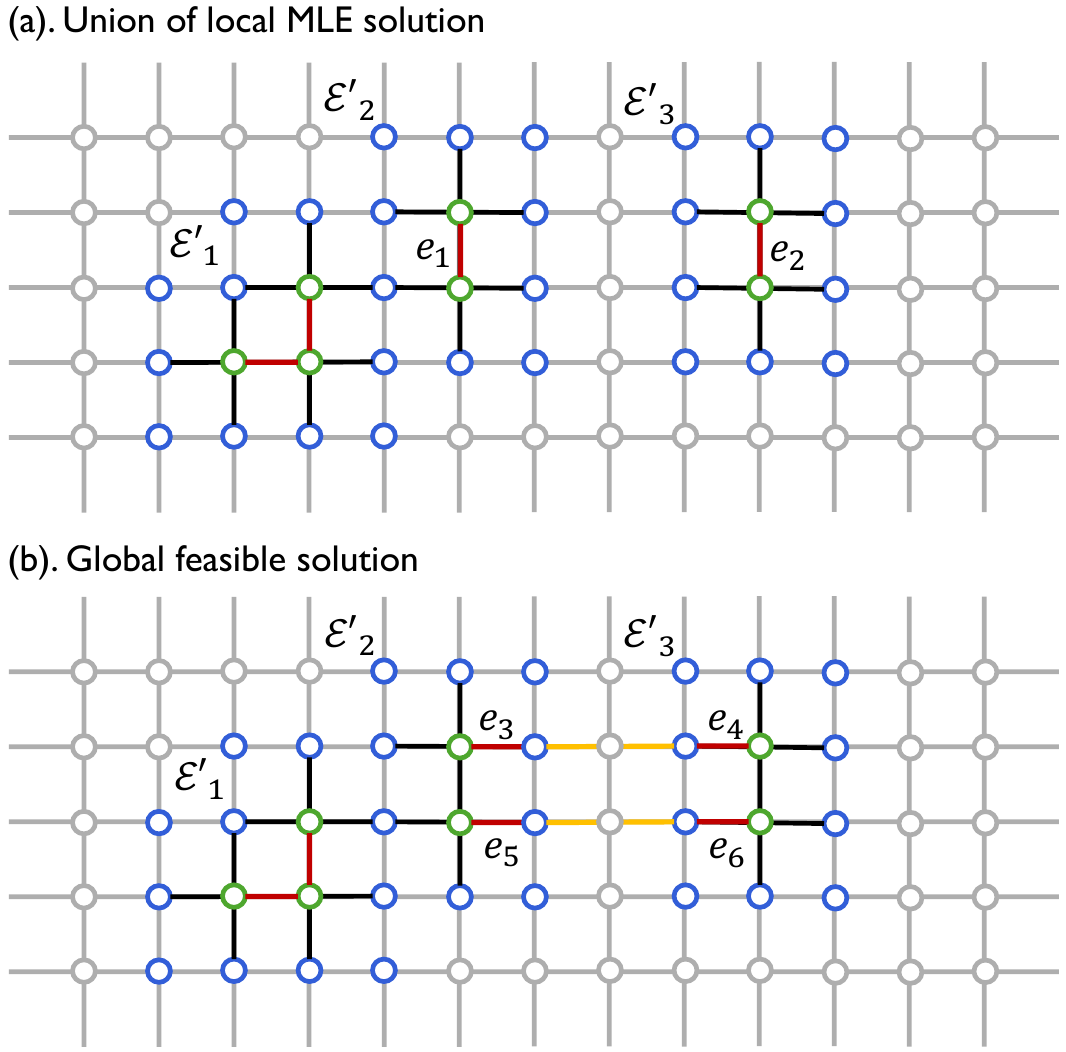}
    \caption{Optimality of fusion decoding.}
    \label{fig:example}
\end{figure}

\begin{lemma}
\label{lem:disjoint}
Under condition~(ii) of Theorem~\ref{thm:fusion}, the interior error sets are pairwise
disjoint: $\mathcal{E}'_i \cap \mathcal{E}'_j = \emptyset$ for all $i \neq j$.
\end{lemma}

\begin{proof}
Suppose $e \in \mathcal{E}'_i \cap \mathcal{E}'_j$ with $i \neq j$. By
definition~\ref{def:cluster}, $e$ is incident to some interior detector $d \in
\mathcal{I}'_i$ and to some interior detector of $\mathcal{D}'_j$. Since $d$ is
interior to $\mathcal{D}'_i$, every error incident to it stays inside that cluster, so
$D(e) \subseteq \mathcal{D}'_i$; the same argument in cluster $j$ gives $D(e) \subseteq
\mathcal{D}'_j$. Hence $D(e) \subseteq \mathcal{D}'_i \cap \mathcal{D}'_j \subseteq
\partial'_i$ by condition~(ii). But $d \in D(e)$ and $d \in \mathcal{I}'_i =
\mathcal{D}'_i \setminus \partial'_i$, a contradiction.
\end{proof}

\begin{proof}[Proof of Theorem~\ref{thm:fusion}]
By condition~(ii), a detector interior to one cluster belongs to no other: if $d \in
\mathcal{I}'_i$ were also in $\mathcal{D}'_j$ for some $j \neq i$, then $d \in
\mathcal{D}'_i \cap \mathcal{D}'_j \subseteq \partial'_i$, contradicting $d \in
\mathcal{I}'_i$. Together with condition~(i), the sets $\mathcal{S}'_i$ therefore
uniquely partition $\mathcal{S}$.

\emph{Feasibility.} Each $\mathcal{F}'^{*}_i$ is boundary-free, so
$\mathcal{F}'^{*}_i \subseteq \mathcal{E}'_i$ and $D(\mathcal{F}'^{*}_i) =
\mathcal{S}'_i$. By Lemma~\ref{lem:disjoint} the union in Equation~\ref{eq:fusion} is
disjoint, so $D(\mathcal{F}^{*}) = \bigoplus_i D(\mathcal{F}'^{*}_i) =
\bigoplus_i \mathcal{S}'_i = \mathcal{S}$, the last step because the $\mathcal{S}'_i$
uniquely partition $\mathcal{S}$. The union is therefore a feasible global solution. Moreover,
each $\mathcal{F}'^{*}_i$ lies in $\mathcal{E}'_i$ and the $\mathcal{E}'_i$ are pairwise
disjoint by Lemma~\ref{lem:disjoint}, so the local solutions are themselves pairwise
disjoint and their weights add, $W(\mathcal{F}^{*}) = \sum_i W(\mathcal{F}'^{*}_i)$.

\emph{Optimality.} Let $\mathcal{F}$ be any feasible global solution, $D(\mathcal{F}) =
\mathcal{S}$, and put $\mathcal{F}_i = \mathcal{F} \cap \mathcal{E}'_i$. Fix an interior
detector $d \in \mathcal{I}'_i$. Every error incident to $d$ lies in $\mathcal{E}'_i$,
so the errors of $\mathcal{F}$ that fire $d$ are exactly those of $\mathcal{F}_i$,
therefore $D(\mathcal{F}_i)\cap \mathcal{I}'_i= \mathcal{S}'_i$. Since $D(\mathcal{F}_i) \subseteq \mathcal{D}'_i$, adding a virtual boundary error for each detector of
$D(\mathcal{F}_i) \cap \partial'_i$ turns $\mathcal{F}_i$ into a feasible solution of
the local problem of cluster $i$ at unchanged weight, whence
$W(\mathcal{F}'^{*}_i) \leq W(\mathcal{F}_i)$ due to the local optimality of $\mathcal{F}'^{*}_i$. The $\mathcal{F}_i$ are disjoint
subsets of $\mathcal{F}$ by Lemma~\ref{lem:disjoint} and weights are non-negative, so:
\begin{equation}
\label{eq:fusion-opt}
W(\mathcal{F}^{*}) \;=\; \sum_{i=1}^{k} W(\mathcal{F}'^{*}_i)
\;\leq\; \sum_{i=1}^{k} W(\mathcal{F}_i)
\;\leq\; W(\mathcal{F}) .
\end{equation}
Thus $\mathcal{F}^{*}$ is feasible and has the smallest weight.
\end{proof}

Figure~\ref{fig:fusion-theory} shows an example. We assume every error mechanism
carries the same weight.
Figure~\ref{fig:fusion-theory}(b) shows the outcome: each cluster is decoded on its own
local DEM, every solution shown in red edges comes out boundary-free, and their
union is the global minimum-weight solution.

Figure~\ref{fig:example} illustrates why no other solution can do better for the same decoding problem in Figure~\ref{fig:fusion-theory}(a). Panel~(a)
shows the union $\mathcal{F}^{*}$ of the local solutions: the local decoding solution of
cluster~2 contains the error $e_1$, and the local decoding solution of cluster~3
contains the error $e_2$. 

Panel~(b) shows a different globally feasible solution
$\mathcal{F}$ for the same syndrome, drawn in both red and yellow edges. Restricting $\mathcal{F}$ to the interior errors of each cluster gives $\mathcal{F}_i =
\mathcal{F} \cap \mathcal{E}'_i$: cluster~2 contributes the pair $\{e_3, e_5\}$ and
cluster~3 contributes the pair $\{e_4, e_6\}$, they are shown in red edges, while the yellow edges indicate errors belong to no $\mathcal{F}_i$
at all. Each pair agrees with the syndrome on its own cluster's interior detectors, so together
with virtual boundary errors it is a feasible solution of that cluster's local problem
and therefore cannot cost less than the local minimum: $e_1$ alone already settles
cluster~2, and $e_2$ alone settles cluster~3. The yellow errors only add further weight.
Summing over the clusters gives Equation~\ref{eq:fusion-opt}.

\section{Real-Time Low-Latency Fusion Decoding}
\label{sec:realtime}

Theorem~\ref{thm:fusion} is a composition rule: if local clusters have
boundary-free solutions and meet only on boundary detectors, the union of their
local answers is a global optimum.  This section gives the online procedure
that reaches this state.  The decoder starts from small fired cores, decodes
their radius-one footprints, grows only the footprints whose solve touches the
virtual boundary, and publishes only when every live footprint is boundary-free.

\subsection{Error-Hop Cluster Footprints and Initialization}
\label{subsec:construction}

The first question is purely geometric: which fired detectors could have been
created by nearby error mechanisms?  We measure this in \emph{error hops}.  One
hop means that two detectors appear together in the same error mechanism.

\begin{definition}[Error-hop distance]
\label{def:error-hop-distance}
Two distinct detectors are one \emph{error hop} apart if they appear in the same
error mechanism.  For a detector set $X\subseteq\mathcal{D}$, the error-hop
distance from $d$ to $X$ is the minimum number of such hops needed to reach any
detector in $X$:
\begin{equation}
\label{eq:dist-set}
\operatorname{dist}(d,X)=
\min\left\{
m \ \middle|\ 
\begin{array}{l}
\exists d_0,\ldots,d_m\in\mathcal{D}: d_0=d,\ d_m\in X,\\
\forall j<m,\ \exists e_j\in\mathcal{E}\ \text{s.t.}\ \{d_j,d_{j+1}\}\subseteq D(e_j)
\end{array}
\right\},
\end{equation}
with the convention that the distance is $\infty$ when $d$ is disconnected from
$X$.
\end{definition}

The fired syndrome is then split into maximal one-hop connected pieces.  These
pieces are not yet decoding regions; they are only the fired seeds from which
regions will be built.

\begin{definition}[Fired core]
\label{def:fired-core}
Let $\mathcal{S}\subseteq\mathcal{D}$ be the fired syndrome.  The \emph{fired
cores} are the connected components of $\mathcal{S}$ under one-hop detector
adjacency.
\end{definition}

To decode a fired core, the decoder must also include nearby unfired detectors:
otherwise an error incident to the core may leave the local region immediately.
The detector set used for local decoding is the cluster footprint.

\begin{definition}[Cluster footprint]
\label{def:radius}
For a cluster with nonempty fired core $C\subseteq\mathcal{S}$ and integer
radius $r\geq 0$, the radius-$r$ \emph{cluster footprint} is:
\begin{equation}
\label{eq:radius-cluster}
B_r(C)=\{d\in\mathcal{D}\mid \operatorname{dist}(d,C)\leq r\}.
\end{equation}
The boundary, interior, and local DEM of a cluster footprint are always
recomputed from the original global DEM using Definitions~\ref{def:cluster}
and~\ref{def:local-dem}.
\end{definition}

Initialization is the radius-one footprint.  After partitioning the observed
syndrome into fired cores $C_1,\ldots,C_m$, the decoder creates one initial
cluster per core:

\[
\mathcal{D}_i=B_1(C_i).
\]
Radius one is the smallest footprint that contains every DEM error mechanism
incident to the fired core.  It also gives the first structural invariant:
different initial footprints can overlap only on boundary detectors.

\begin{proposition}[Initialized footprints overlap only on boundary]
\label{prop:init-invariant}
For any two distinct fired cores $C_i$ and $C_j$, their initial cluster
footprints satisfy:
\[
B_1(C_i)\cap B_1(C_j)
\subseteq
\partial B_1(C_i)\cap \partial B_1(C_j).
\]
\end{proposition}

\begin{proof}
Distinct fired cores are distinct connected components of the fired syndrome, so
no fired detector in $C_j$ is within one error hop of $C_i$.  Suppose some
$d\in B_1(C_i)\cap B_1(C_j)$ were interior to $B_1(C_i)$.  Since
$d\in B_1(C_j)$, there is an $s_j\in C_j$ with
$\operatorname{dist}(d,s_j)\leq 1$.  If $d=s_j$, then $s_j$ is already within
one hop of $C_i$.  Otherwise, $d$ and $s_j$ share an error mechanism; because
$d$ is interior to $B_1(C_i)$, that mechanism is contained in $B_1(C_i)$, so
$s_j$ is again within one hop of $C_i$.  Both cases contradict the separation of
the fired cores.  Hence the overlap contains no interior detector of
$B_1(C_i)$, and by symmetry none of $B_1(C_j)$.
\end{proof}

\begin{figure}[t]
    \centering
    \includegraphics[width=0.7\linewidth]{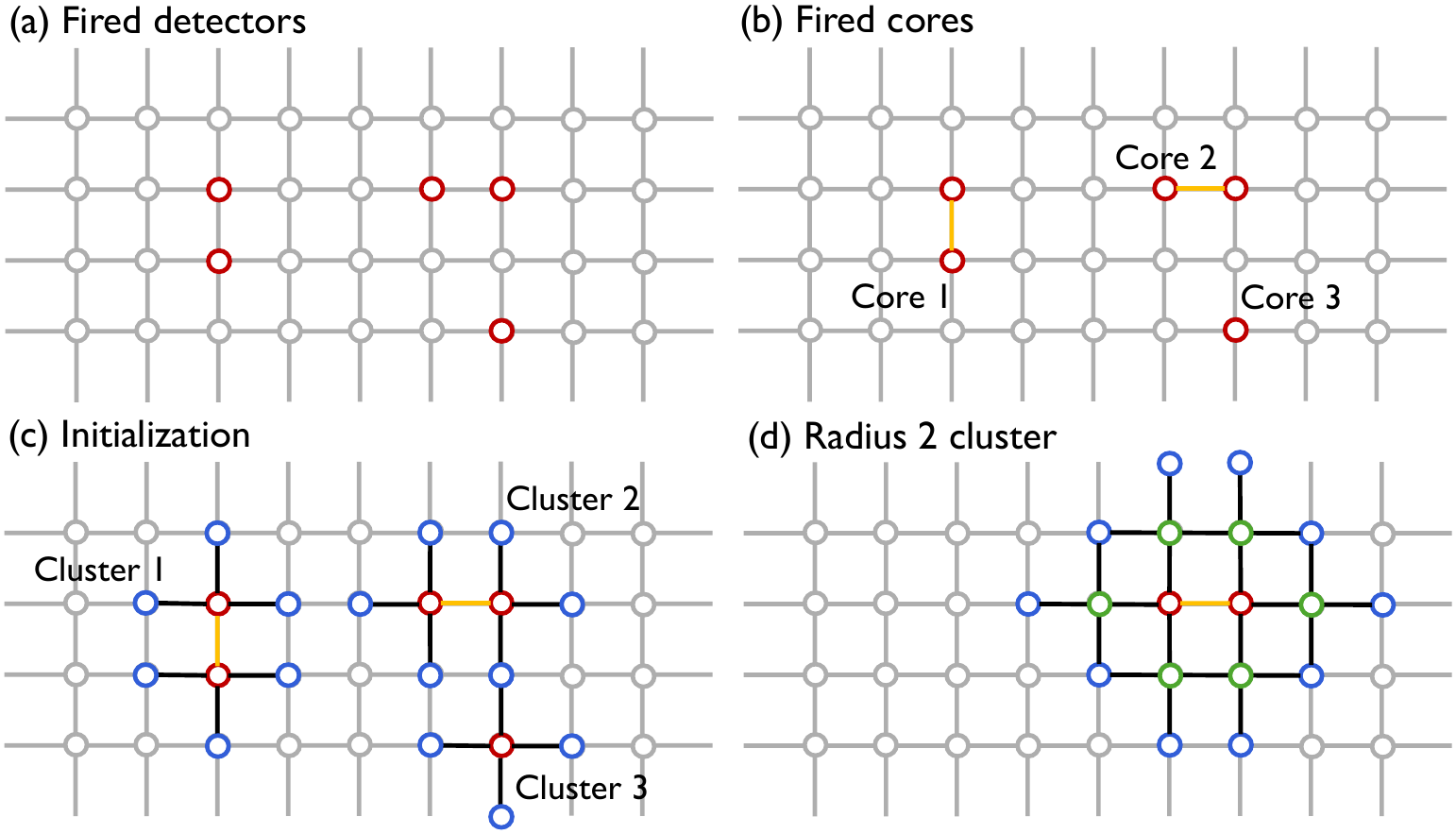}
    \caption{Radius-one cluster-footprint initialization.}
    \label{fig:init}
\end{figure}

Figure~\ref{fig:init} shows the construction.  Fired detectors are first grouped
into fired cores, where we use yellow edges to represent one error hop connections, then each core receives its radius-one footprint.  Red
detectors belong to fired cores; blue detectors are boundary detectors added by
the DEM neighborhood.  Panel (d) shows the same idea at radius two: growing the
footprint converts some former boundary detectors into interior detectors
(green), and the black edges are the error mechanisms now fully internal to the
cluster.

\begin{remark}[Why start from radius one]
Radius one is also the useful practical default.  In low-error regimes, most
nontrivial fired cores are explained by one local error mechanism; cores that
require two, three, or more nearby mechanisms appear with rapidly decreasing
frequency.  The decoder therefore solves the dominant case immediately and grows
only when the local solve says that the radius-one footprint is not enough.
\end{remark}

\subsection{Cluster Growth and Union}
\label{subsec:resolve}

Initialization gives the smallest local problems; the first local solve decides
which of them are already large enough.  A boundary-free solution is kept, but
only tentatively: a neighboring cluster may later grow into this footprint and
change the local problem.  A boundary-dependent solution is the grow signal.  It
means that the current footprint can explain the syndrome only by using a
virtual boundary error.

Growth is anchored at the original fired core.  If a cluster began at $C_i$ and
has radius $r_i$, the next grow step replaces its grown part by
$B_{r_i+1}(C_i)$.  Detectors absorbed from other clusters do not become new
growth centers.  They enter only through cluster union.

\begin{definition}[Cluster interaction]
\label{def:cluster-interaction}
For two active clusters with recomputed cluster footprints
$U,V\subseteq\mathcal{D}$, say that $U$ reaches $V$ if their overlap is not purely boundary-boundary:
\[
U\cap V \not\subseteq \partial(U)\cap\partial(V).
\]
Equivalently, since each detector in a footprint is either interior or boundary,
\[
(\partial(U)\cap\mathcal{I}(V))
\cup
(\partial(V)\cap\mathcal{I}(U))
\cup
(\mathcal{I}(U)\cap\mathcal{I}(V))
\neq\emptyset .
\]
Here $\partial(U)$, $\mathcal{I}(U)$, $\partial(V)$, and $\mathcal{I}(V)$ are
computed in the global DEM.  This is the only trigger for cluster union: when
two active footprints have an overlap that would violate the boundary-only
intersection condition, the two clusters are combined.
\end{definition}

Cluster union is literal detector-set union.  If $U$ reaches active footprints
$V_1,\ldots,V_q$, the combined footprint is:
\[
U^+=U\cup V_1\cup\cdots\cup V_q .
\]

After union, the decoder recomputes the boundary, interior, and local DEM of
$U^+$ from the global DEM.  It does not glue old local DEMs together: an error
mechanism that previously escaped through a boundary may now be fully internal.

Union is closed before solving again. 
After each union, the decoder checks the new combined footprint against the
other active footprints.  If it reaches another active footprint, that footprint
is unioned as well.
The next local solve starts only after this closure stops. Equivalently,
closure repeatedly unions every pair of live footprints whose overlap is not
purely boundary-boundary. At that point, every remaining intersection between
active footprints is boundary-only, exactly the intersection condition needed by
Theorem~\ref{thm:fusion}.

\begin{figure}[t]
    \centering
    \includegraphics[width=0.7\linewidth]{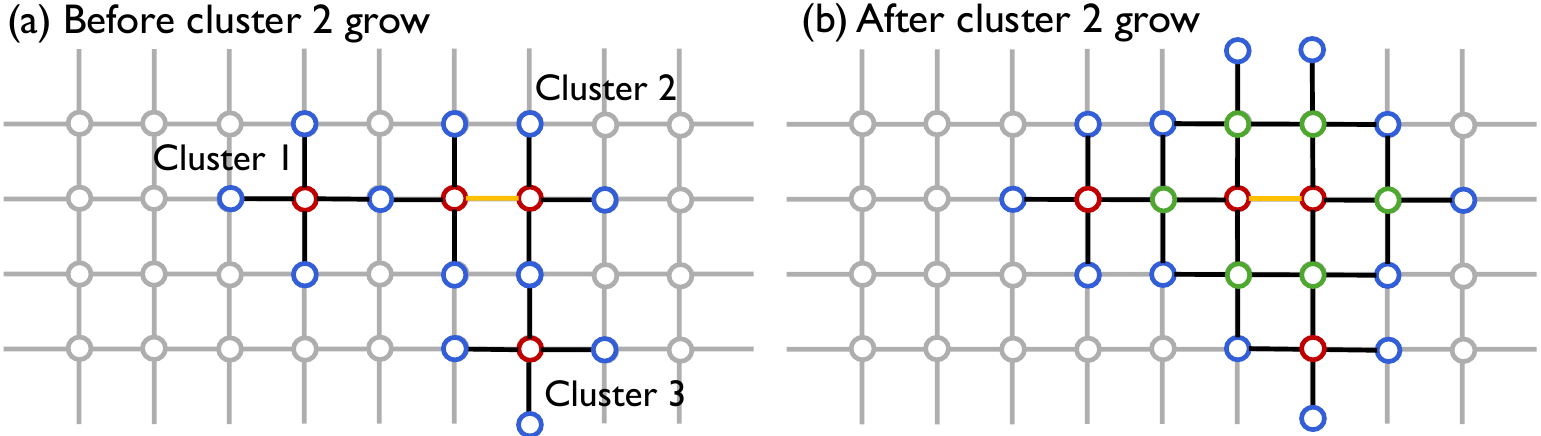}
    \caption{Example of cluster growth and union.}
    \label{fig:cluster-union-example}
\end{figure}

Figure~\ref{fig:cluster-union-example} shows one closure step.  In Panel (a),
the three radius-one footprints are active separately.  Suppose Cluster 2 is
boundary-dependent.  The decoder grows Cluster 2 by an additional one error hop from its
original cluster to radius 2.  In Panel (b), the grown boundary reaches the interiors of
both Cluster 1 and Cluster 3, so all three footprints are replaced by their
detector-set union.  The combined boundary, interior, and local DEM are then
recomputed before the next local solve.

\subsection{Online Cluster Lifecycle}
\label{subsec:scheduler}
\label{subsec:algorithm}

The geometry above is static; the implementation is streaming.  Detector
outcomes arrive round by round, but the background DEM is known from the start.
The scheduler therefore uses the DEM to plan growth and union early, and waits
for detector outcomes only when it is time to initialize, construct a local DEM,
or launch a solve.  The easiest way to state the rule is to follow one cluster
through its lifecycle.

At QEC round $t$, let
$\mathcal{A}_t\subseteq\mathcal{D}$ be the detectors whose outcomes have arrived,
and let $\mathcal{S}_t\subseteq\mathcal{A}_t$ be the fired detectors observed so
far.  For a detector set $U$, write
\[
\mathsf{ready}_t(U)\iff U\subseteq\mathcal{A}_t .
\]

There are two gates.  The \emph{data gate} is $\mathsf{ready}_t(U)$: the
scheduler cannot build or solve a local DEM until all detectors in its footprint
have arrived.  The \emph{core gate} applies during growth: if a grow footprint
contains a fired detector that still belongs to a provisional fired core, the
growing cluster waits until that core finalizes and initializes its own
radius-one footprint.  Only then can ordinary cluster union absorb it.

\textbf{Birth: provisional fired core.}
A newly fired detector first joins any provisional fired core within one error
hop; if none exists, it starts a new provisional core.  The scheduler then checks
the one-hop frontier of that core.  If part of the frontier lies in future QEC
rounds, the core waits.  When those outcomes arrive, any fired frontier detector
is merged into the core, and the frontier check repeats.  The fired core $C$ is
finalized exactly when the radius-one frontier has arrived and adds no new fired
detector:
\[
\mathsf{ready}_t(B_1(C))
\qquad\text{and}\qquad
B_1(C)\cap\mathcal{S}_t\subseteq C .
\]

\textbf{Initialization: owner task.}
Once $C$ is finalized, it becomes an initialized cluster task with
footprint $B_1(C)$.

\begin{definition}[Cluster task]
\label{def:cluster-task}
A \emph{cluster task} is either an \emph{owner} or \emph{masked}.  An owner
controls one active footprint.  A masked task has been absorbed by another owner
and is no longer scheduled.
An owner task $K$ stores:
\[
K=(C_K,r_K,M_K,\mathcal{D}_K,\sigma_K).
\]
Here $C_K$ is the original fired core, $r_K$ is the radius grown around
that core, $M_K$ is the set of finalized fired cores currently owned by $K$,
and $\mathcal{D}_K$ is the current union footprint.  Initially,
\[
C_K=C,\qquad r_K=1,\qquad M_K=\{C\},\qquad \mathcal{D}_K=B_1(C).
\]
$\partial_K$ and $\mathcal{I}_K$ are derived from $\mathcal{D}_K$ and are
recomputed in the global DEM whenever the footprint changes.

The task status is
\[
\sigma_K\in
\{\mathsf{Waiting},\mathsf{Decoding},
  \mathsf{BoundaryFree},\mathsf{BoundaryDependent}\}.
\]
$\mathsf{Waiting}$ means that an outcome or a dependent provisional core is
missing.  $\mathsf{Decoding}$ means that a local solve is in flight.
$\mathsf{BoundaryFree}$ and $\mathsf{BoundaryDependent}$ record the most recent
decoding solve.  A boundary-free answer is tentative until the global publish
condition holds.
\end{definition}

Ownership is a scheduling device, not a publication rule.  It prevents two
clusters from doing duplicate growth after one has already absorbed the other,
and it makes stale local-solve results easy to discard.

\textbf{First solve.}
When $\mathsf{ready}_t(\mathcal{D}_K)$ holds, $K$ solves its radius-one local
DEM.  A boundary-free result is stored as a tentative answer.  A
boundary-dependent result sends $K$ into the grow--union--solve loop.

\textbf{Grow--union--solve loop.}
Suppose owner $K$ is boundary-dependent.  One loop iteration consists of four
steps.

\begin{figure}[t]
    \centering
    \includegraphics[width=0.9\linewidth]{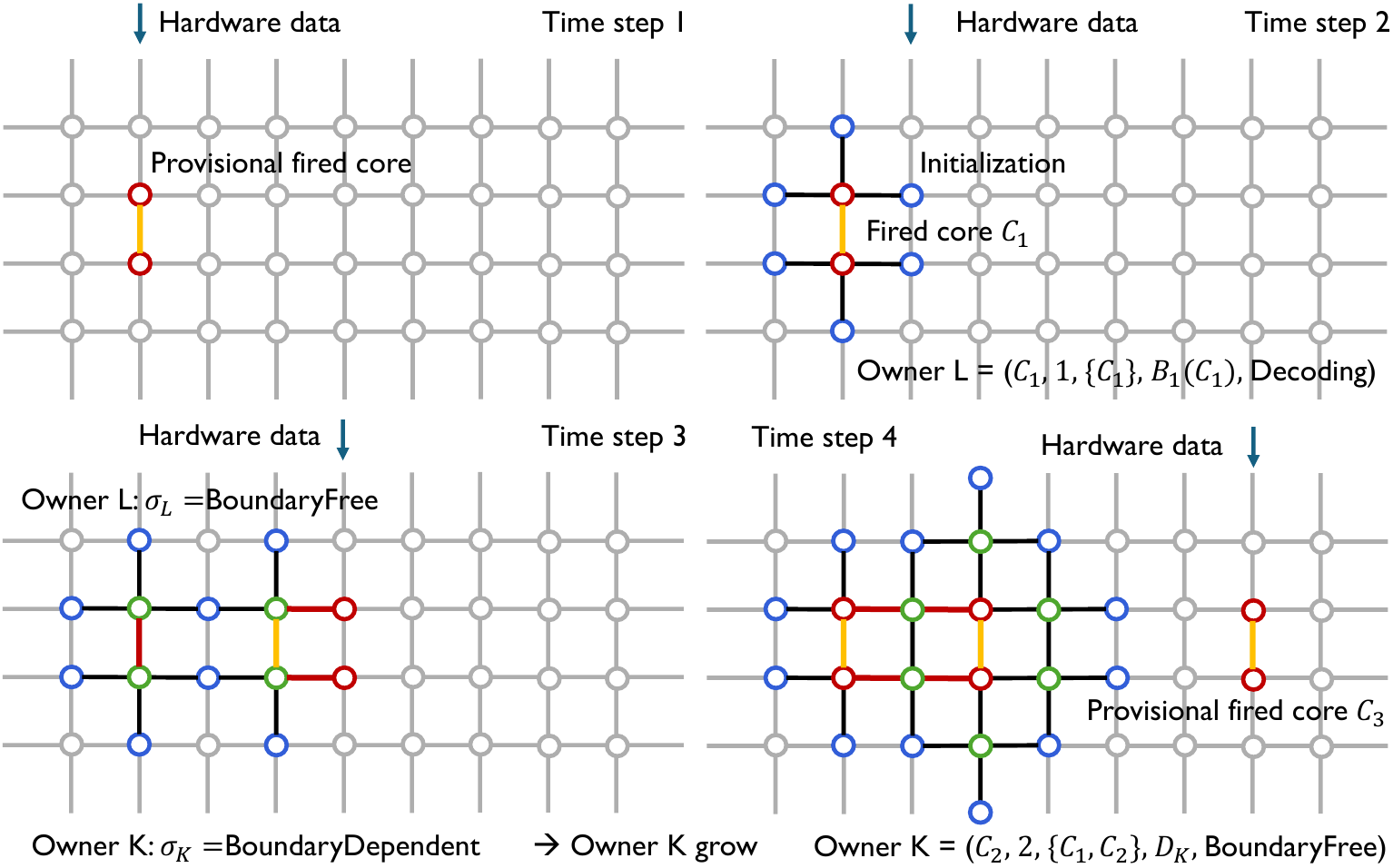}
    \caption{Online lifecycle.}
    \label{fig:online-lifecycle-example}
\end{figure}

\begin{enumerate}[leftmargin=*]
\item \textbf{Grow the anchored footprint.}
The radius grows only around the original fired core.  The initial grow
candidate is
\[
U_0
=
\mathcal{D}_K \cup B_{r_K+1}(C_K).
\]
Only the anchored ball $B_{r_K+1}(C_K)$ is enlarged; the already claimed
footprints contained in $\mathcal{D}_K$ are carried forward by union, not grown.

\item \textbf{Plan cluster union in the background DEM.}
Starting from $U_0$, the scheduler scans for reached owners.  A live owner
$L\neq K$ is reached when the current planned footprint $U$ interacts with
$\mathcal{D}_L$ in the sense of Definition~\ref{def:cluster-interaction};
equivalently, their overlap is not purely boundary-boundary.  Reaching
an owner means claiming all fired cores currently owned by that owner, not
necessarily touching those fired cores themselves.
If those cores are already in $M_K$, nothing new is claimed.  Otherwise, $K$
claims $L$: the task $L$ is masked, its tentative answer is withdrawn, and its
owned cores are added to $M_K$.  The planned footprint is updated by detector
union,
\[
U\gets U\cup\mathcal{D}_L,
\]
and the scan repeats, because the enlarged union footprint may interact with
additional owner footprints.

\item \textbf{Wait for unresolved data or cores.}
The same geometric scan also checks whether $U$ contains a fired detector whose
fired core is still provisional.  Such a provisional core cannot be absorbed as
an isolated detector.  It is recorded as a dependency, and $K$ waits until the
core finalizes and initializes its own radius-one footprint.  The scheduler then
reruns the same geometric closure at the same radius, now treating the newly
initialized core as an ordinary owner that may be claimed if its footprint
interacts with $U$.

The plan can be committed only after two gates clear: $\mathsf{ready}_t(U)$ holds for the final planned footprint, and no reached fired detector belongs to
an unresolved provisional core.  If detector outcomes are still missing, $K$
stores the current grow candidate and waits; when the outcomes arrive, it
resumes the same radius rather than growing again.

\item \textbf{Commit and solve.}
Once the gates clear, $K$ commits the union: $\mathcal{D}_K$ becomes the final
union footprint, $\partial_K$ and $\mathcal{I}_K$ are recomputed in the global
DEM, the new local DEM is built, and the next local solve is launched.  The loop
ends only when $K$ becomes boundary-free or is masked by another owner.
\end{enumerate}

Figure~\ref{fig:online-lifecycle-example} illustrates the online lifecycle.  At
time step 1, the hardware has revealed only enough data to identify a
provisional fired core, so the scheduler holds it while waiting for its one-hop
frontier.  At time step 2, the frontier has arrived and no new fired detector
extends the core; the scheduler finalizes it as fired core $C_1$, initializes
the radius-one footprint $B_1(C_1)$, and launches its first local solve as owner
$L$.  At time step 3, owner $L$ has a boundary-free tentative answer, but this
answer is not publishable because another live owner $K$ is still
boundary-dependent.  Owner $K$ therefore grows from its own original fired core
$C_2$.  At time step 4, the grown footprint of $K$ interacts with the footprint
owned by $L$, so $K$ claims $L$: the task $L$ is masked, the core $C_1$ is added
to $M_K$, and the new owner footprint becomes the detector union of the two
footprints.  The scheduler then recomputes the boundary and interior of the
combined footprint and launches the next local solve for $K$. In the meantime, new provisional fired core $C_3$ are tracked.

\textbf{Interruptions.}
A task can pause for missing detector outcomes or for a provisional fired core
that it has reached but cannot yet absorb.  It can also terminate without
publishing: if another owner reaches its interior first, this task is masked,
its tentative answer is withdrawn, and any in-flight solve result is stale.

\textbf{Global publication.}
The decoder publishes only at an all-clear fixed point:
\[
\mathsf{Publish}
\iff
\left\{
\begin{array}{l}
\text{no provisional fired core remains,}\\
\text{no live owner is waiting, decoding, or boundary-dependent,}\\
\text{every live owner has a tentative boundary-free answer.}
\end{array}
\right.
\]
The published correction is
\[
\widehat{E}
=
\bigcup_{K\ {\rm live}} A_K^{\rm real},
\]
where $A_K^{\rm real}$ is the real-error part of $K$'s tentative local answer.
At this fixed point, the live footprints cover the observed syndrome, all masked
tasks have no output, closure has restored boundary-only intersections, and
every live local answer is boundary-free.  Theorem~\ref{thm:fusion} therefore
applies directly.  With a minimum-weight local backend, $\widehat{E}$ is a
global minimum-weight solution.

\begin{remark}[Why online fusion reduces latency]
The scheduler does not wait for the full syndrome volume before starting local
decoding.  As soon as a fired core is finalized and its cluster footprint is
ready, its local solve can run while later QEC rounds are still being acquired.
Most clusters become boundary-free at small radius and therefore finish before
the final measurement round arrives.  The post-acquisition critical path is then
only the unfinished tail: unresolved provisional cores, pending grow steps,
remaining local solves, and final publication.  This overlap between syndrome
acquisition and local decoding is the source of the latency reduction measured
in Section~\ref{sec:evaluation}.
\end{remark}

\subsection{Solver Interface, Fallbacks, and Latency}
\label{subsec:solver-interface}

A key design point of \myCompilerNameSpace is that the scheduler is
\textbf{backend-agnostic}.  It does not assume a particular decoding algorithm,
graph structure, or solver implementation.  The only requirement is that the
backend can solve the local DEM and report whether its selected solution used
the virtual boundary.

The scheduler needs only a narrow local-decoder interface.  For a cluster
footprint $\mathcal{D}'$, let $\mathcal{I}'$ be its recomputed interior.  The
assigned fired syndrome lies in $\mathcal{I}'$, so the backend is called as:
\[
\mathsf{Solve}(\mathsf{LocalDEM}(\mathcal{D}'),\mathcal{S}\cap\mathcal{I}')
\longrightarrow
(\widehat E', b') .
\]
Here $\widehat E'$ is the real-error set selected by the backend, and $b'$
indicates whether any virtual boundary error was used.  If $b'=0$, the scheduler
stores $\widehat E'$ as a tentative boundary-free answer.  If $b'=1$, the
cluster is boundary-dependent and must grow.  With an exact backend, the publish
fixed point gives the MLE guarantee of Theorem~\ref{thm:fusion}; with a
heuristic backend, the same schedule is used but optimality is empirical.

This narrow interface is what makes fusion decoding generic.  The local solver
may be an exact MLE solver, a matching-style decoder, a belief-propagation based
heuristic, or a hardware-accelerated local optimizer.  These choices affect the
quality and runtime of each local solve, but not the scheduler logic: grow,
cluster union, masking, and publication depend only on $\widehat E'$ and $b'$.

The implementation also runs a monolithic decoder in parallel as a conservative
fallback.  This matters only for rare tail events.  For example, a long error
chain can make one cluster grow many times, producing a large-radius footprint
whose local problem is no longer much smaller than the monolithic one.  The two
paths therefore race: the first valid answer to finish is used.  At low physical
error rate, such long-chain events are exponentially suppressed in the chain
length.

Now we define the latency for fusion decoding.

\begin{definition}[Post-acquisition decoding latency]
\label{def:stages}
Let $T_{\mathsf{acq}}$ be the time when the final syndrome-measurement round has
arrived, and let $T_{\mathsf{pub}}^{\mathsf{fusion}}$ be the time when the
fusion scheduler reaches the $\mathsf{Publish}$ fixed point of
Section~\ref{subsec:algorithm}.  The post-acquisition fusion latency is:
\[
\Delta_{\mathsf{fusion}}
=
T_{\mathsf{pub}}^{\mathsf{fusion}}-T_{\mathsf{acq}} .
\]
\end{definition}

\section{Evaluation}
\label{sec:evaluation}

We evaluate whether \myCompilerNameSpace converts decoding from an end-of-acquisition bottleneck into computation that proceeds concurrently with syndrome acquisition.
Our experiments therefore measure the post-acquisition decoding latency defined in Definition~\ref{def:stages}, compare the resulting real-time critical path with monolithic and parallel-window baselines, and evaluate its scaling with spatial workload, circuit depth, and physical error rate.

\subsection{Experimental Setup}
\label{subsec:experiments-setup}

\textbf{Baselines.} We compare \myCompilerNameSpace against two baselines. The first is
\emph{monolithic decoding}, which waits for the complete syndrome volume to be acquired before invoking a decoder once on the entire detector error model (DEM). The second is
\emph{parallel window decoding}~\cite{skoric2023parallel}, which partitions the DEM along the
time axis into fixed-thickness $3d$-round slabs, decoded in two interleaved layers. For both
baselines and \myCompilerName, we instantiate the \emph{same} inner decoder -- Minimum-Weight
Perfect Matching (MWPM)~\cite{higgott2022pymatching} for graphlike surface code DEMs, the
search-based Tesseract decoder~\cite{beni2025tesseract} for general DEMs, or the
heuristic relay-BP decoder~\cite{muller2025improved} -- so that every latency comparison isolates
the effect of the decoding \emph{strategy} rather than the underlying solver.

\textbf{Experimental Configurations.}
The overall study compares representative code--backend pairs at a common operating point; the scalability study independently varies code-surgery width, code distance, and circuit depth; the window study compares FusionQ with the parallel-window-decoding baseline over long syndrome histories; and the physical-error sweep reveals a new, code- and backend-dependent fusion-decoding threshold.
Unless stated otherwise, we assume a syndrome-acquisition period of
$\tau = 5\,\mathrm{ms}$, representative of the QEC-cycle times reported
for neutral-atom hardware and projected for trapped-ion
architectures~\cite{bluvstein2026fault,loschnauer2025scalable,hughes2025trapped}.
Each subsection explicitly specifies the QEC depth used in its experiments.

\textbf{Benchmarks.}
We evaluate three structurally different code families: surface codes $\llbracket d^2,1,d\rrbracket$ with $d\in\{5,7,9,11\}$; BB codes~\cite{bravyi2024high} with parameters $\llbracket72,12,6\rrbracket$, $\llbracket90,8,10\rrbracket$, $\llbracket108,8,10\rrbracket$, and $\llbracket144,12,12\rrbracket$; and an HGP code $\llbracket420,10,10\rrbracket$~\cite{aydin2025cyclic}.
For the code-surgery study, GeneCS~\cite{zhou2026genecs} synthesizes joint logical-measurement circuits over multiple BB-code blocks.
All circuits use the SI1000 circuit-level noise model~\cite{gidney2022benchmarking}.

\textbf{Metrics.} We report four metrics, drawn from per-shot decoding traces.
(\textbf{a}) \textbf{Post-acquisition latency}:
the interval from the final syndrome arrival to publication of a correction, $t_{\mathrm{publish}}-t_{\mathrm{acq}}$, reported in milliseconds.
This metric directly measures the decoder contribution to the real-time feedback deadline: computation completed before $t_{\mathrm{acq}}$ is hidden by data acquisition.
(\textbf{b}) \textbf{Logical error rate (LER)}:
the observed logical-failure count normalized by the number of shots and QEC rounds, $N_{\mathrm{fail}}/(N_{\mathrm{shots}}L)$, for the $200$K-shot overall study; smaller studies report raw failure counts.
(\textbf{c}) \textbf{Cost consistency}: on every shot, we compare \myCompilerName's solution
cost directly against the baseline cost and classify it as strictly \emph{lower} (\textbf{L}) or
strictly \emph{higher} (\textbf{H}); the remainder tie. MWPM never mismatches; Tesseract's search is
capped by \texttt{pqlimit}, so it is exact only in principle and can rarely mismatch. We report L/H
for both exact backends as raw shot counts; for the heuristic relay-BP backend, which has no
exact-minimum contract, we report L/H as percentages of shots.
(\textbf{d}) \textbf{Speedup}:
the ratio of baseline to \myCompilerNameSpace post-acquisition latency, computed separately for the mean and each reported percentile.

\textbf{Implementation.} (\textbf{a}) \textit{Decoders.}
All three backend decoders: MWPM, Tesseract, and Relay-BP are implemented in Rust, we set \texttt{pqlimit} of Tesseract to $10^7$. Fusion, monolithic, and parallel window decoding always call the same decoder of each experiment.
We implement the FusionQ runtime and all three backend interfaces in Rust.
Tesseract uses a priority-queue limit of $10^7$ in the overall, scalability and long-window studies, and $10^8$ for the BB code physical-error sweep in Section~\ref{subsec:fusion-threshold}.
(\textbf{b}) \textit{Latency.} Every job uses one CPU core for a
fair comparison across strategies.
We measure each local solve on that core and replay its recorded readiness and duration under an ideal parallel-resource model: independent tasks may overlap, whereas re-decodes of the fusion cluster and the causal $A$-to-$B$ window dependencies of window decoding remain serialized.
This trace-driven procedure evaluates the scheduler's real-time critical path without allowing concurrent benchmark processes to contaminate wall-clock measurements.
(\textbf{c}) \textit{Machine.} All experiments are submitted one at a time, conducted on a server with a 2.4\,GHz CPU and 0.75\,TB of memory.

\subsection{Overall Results}
\label{subsec:overall-results}

We first ask whether FusionQ can keep pace with a syndrome stream that produces one new round every $5$ms.
We evaluate five representative code--backend pairs in both the $Z$ and $X$ memory bases, as summarized in Table~\ref{tab:overall-configs}.
Every configuration uses $p=5\times10^{-4}$, $200$K shots per basis, and a $5$ms syndrome-production interval.
We set the circuit depth to $L=3d$ and use the full latency distribution to determine, for each configuration, the highest reported percentile whose post-acquisition decoding latency remains below the $5$ms syndrome-production interval.

\begin{table}[t]
\centering
\caption{Experimental configurations for the overall comparison.}
\label{tab:overall-configs}
\small
\setlength{\tabcolsep}{5pt}
\renewcommand{\arraystretch}{1.08}
\resizebox{\linewidth}{!}{%
\begin{tabular}{llcccccc}
\toprule
\textbf{Benchmark} &
\textbf{Code} &
\textbf{Rounds} &
\textbf{Backend} &
\textbf{Physical error} &
\textbf{Shots} &
\textbf{Acq./round} &
\textbf{Cost metric} \\
\midrule
Surface code & $\llbracket 121,1,11\rrbracket$ & $L=33$ & MWPM & $5\times10^{-4}$ & $200$K & $5$ms & L/H (count) \\
BB code & $\llbracket 144,12,12\rrbracket$ & $L=36$ & Tesseract & $5\times10^{-4}$ & $200$K & $5$ms & L/H (count) \\
BB code & $\llbracket 144,12,12\rrbracket$ & $L=36$ & relay-BP & $5\times10^{-4}$ & $200$K & $5$ms & L/H (\%) \\
HGP code & $\llbracket 420,10,10\rrbracket$ & $L=30$ & Tesseract & $5\times10^{-4}$ & $200$K & $5$ms & L/H (count) \\
HGP code & $\llbracket 420,10,10\rrbracket$ & $L=30$ & relay-BP & $5\times10^{-4}$ & $200$K & $5$ms & L/H (\%) \\
\bottomrule
\end{tabular}}
\end{table}

\begin{table*}[t]
\centering
\caption{Post-acquisition latency (ms), per-round logical-failure frequency, and cost-consistency results
for the overall comparison. Each configuration is reported for both the $Z$ and $X$ memory
bases. For zero observed failures, the LER entry reports the observed value and an exact one-sided $95\%$ upper confidence bound, $(1-0.05^{1/N_{\mathrm{shots}}})/L$, under independent-shot sampling.}
\label{tab:overall-latency}
\scriptsize
\setlength{\tabcolsep}{2.2pt}
\renewcommand{\arraystretch}{1.1}
\resizebox{\textwidth}{!}{%
\begin{tabular}{|c|c|c|c|c|c|c|c|c|c|c|c|c|c|c|c|c|c|}
\hline
\textbf{Configuration} &
\textbf{Solver} &
\textbf{Basis} &
\textbf{Report} &
\textbf{mean} &
\textbf{p20} &
\textbf{p30} &
\textbf{p40} &
\textbf{p50} &
\textbf{p60} &
\textbf{p70} &
\textbf{p80} &
\textbf{p90} &
\textbf{p95} &
\textbf{p99} &
\textbf{worst} &
\textbf{LER} &
\textbf{Cost check} \\
\hline
\multirow{6}{*}{\makecell[l]{Surface code\\$\llbracket 121,1,11\rrbracket$\\$L=33$ rounds}}
& \multirow{6}{*}{MWPM}
& \multirow{3}{*}{$Z$}
& Monolithic & 56.7 & 45.5 & 49.1 & 52.7 & 55.7 & 59.2 & 62.9 & 67.2 & 73.6 & 79.1 & 90.7 & 142.0 & $0\,(<4.6\times10^{-7})$ & reference \\
\cline{4-18}
& & & \myCompilerName & 0.12 & 0.06 & 0.08 & 0.09 & 0.10 & 0.10 & 0.12 & 0.15 & 0.19 & 0.24 & 0.54 & 26.5 & $0\,(<4.6\times10^{-7})$ & 0/0 \\
\cline{4-18}
& & & Speedup & 490$\times$ & 728$\times$ & 619$\times$ & 589$\times$ & 576$\times$ & 566$\times$ & 540$\times$ & 455$\times$ & 394$\times$ & 336$\times$ & 168$\times$ & 5$\times$ & -- & -- \\
\cline{3-18}
& & \multirow{3}{*}{$X$}
& Monolithic & 56.6 & 45.5 & 49.2 & 52.7 & 55.6 & 59.2 & 62.8 & 67.2 & 73.5 & 79.0 & 90.4 & 130.4 & $0\,(<4.6\times10^{-7})$ & reference \\
\cline{4-18}
& & & \myCompilerName & 0.12 & 0.06 & 0.08 & 0.09 & 0.10 & 0.10 & 0.12 & 0.15 & 0.19 & 0.24 & 0.54 & 23.9 & $0\,(<4.6\times10^{-7})$ & 0/0 \\
\cline{4-18}
& & & Speedup & 489$\times$ & 733$\times$ & 614$\times$ & 586$\times$ & 575$\times$ & 564$\times$ & 543$\times$ & 452$\times$ & 394$\times$ & 335$\times$ & 167$\times$ & 6$\times$ & -- & -- \\
\hline
\multirow{6}{*}{\makecell[l]{BB code\\$\llbracket 144,12,12\rrbracket$\\$L=36$ rounds}}
& \multirow{6}{*}{Tesseract}
& \multirow{3}{*}{$Z$}
& Monolithic & 3.3 & 1.2 & 1.3 & 1.3 & 1.4 & 1.5 & 1.7 & 2.0 & 3.3 & 6.0 & 28.9 & 10036.4 & $1.4\times10^{-7}$ & reference \\
\cline{4-18}
& & & \myCompilerName & 0.7 & 0.1 & 0.1 & 0.2 & 0.2 & 0.3 & 0.4 & 0.5 & 0.8 & 1.4 & 5.0 & 9421.8 & $1.4\times10^{-7}$ & 0/0 \\
\cline{4-18}
& & & Speedup & 4$\times$ & 10$\times$ & 9$\times$ & 7$\times$ & 6$\times$ & 5$\times$ & 4$\times$ & 4$\times$ & 4$\times$ & 4$\times$ & 6$\times$ & 1$\times$ & -- & -- \\
\cline{3-18}
& & \multirow{3}{*}{$X$}
& Monolithic & 3.3 & 1.2 & 1.2 & 1.3 & 1.4 & 1.5 & 1.7 & 2.0 & 3.2 & 5.9 & 28.9 & 10476.1 & $4.2\times10^{-7}$ & reference \\
\cline{4-18}
& & & \myCompilerName & 0.7 & 0.1 & 0.1 & 0.2 & 0.2 & 0.3 & 0.4 & 0.5 & 0.9 & 1.5 & 5.3 & 10476.1 & $1.4\times10^{-7}$ & 2/0 \\
\cline{4-18}
& & & Speedup & 4$\times$ & 10$\times$ & 9$\times$ & 7$\times$ & 6$\times$ & 5$\times$ & 4$\times$ & 4$\times$ & 4$\times$ & 4$\times$ & 5$\times$ & 1$\times$ & -- & -- \\
\hline
\multirow{6}{*}{\makecell[l]{BB code\\$\llbracket 144,12,12\rrbracket$\\$L=36$ rounds}}
& \multirow{6}{*}{relay-BP}
& \multirow{3}{*}{$Z$}
& Monolithic & 12.4 & 7.3 & 8.7 & 9.5 & 10.5 & 11.7 & 13.1 & 15.1 & 18.6 & 26.5 & 46.4 & 895.5 & $2.8\times10^{-7}$ & reference \\
\cline{4-18}
& & & \myCompilerName & 2.3 & 0.4 & 0.5 & 0.8 & 1.1 & 1.5 & 2.1 & 3.3 & 5.9 & 8.7 & 15.1 & 412.2 & $1.4\times10^{-7}$ & 45.91\%/16.99\% \\
\cline{4-18}
& & & Speedup & 5$\times$ & 20$\times$ & 16$\times$ & 12$\times$ & 10$\times$ & 8$\times$ & 6$\times$ & 4$\times$ & 3$\times$ & 3$\times$ & 3$\times$ & 2$\times$ & -- & -- \\
\cline{3-18}
& & \multirow{3}{*}{$X$}
& Monolithic & 12.3 & 7.3 & 8.6 & 9.5 & 10.5 & 11.7 & 13.1 & 15.0 & 18.5 & 26.0 & 45.7 & 886.9 & $0\,(<4.2\times10^{-7})$ & reference \\
\cline{4-18}
& & & \myCompilerName & 2.3 & 0.3 & 0.5 & 0.7 & 1.0 & 1.4 & 2.1 & 3.4 & 6.0 & 8.8 & 15.1 & 886.9 & $0\,(<4.2\times10^{-7})$ & 45.75\%/16.83\% \\
\cline{4-18}
& & & Speedup & 5$\times$ & 23$\times$ & 18$\times$ & 14$\times$ & 11$\times$ & 8$\times$ & 6$\times$ & 4$\times$ & 3$\times$ & 3$\times$ & 3$\times$ & 1$\times$ & -- & -- \\
\hline
\multirow{6}{*}{\makecell[l]{HGP code\\$\llbracket 420,10,10\rrbracket$\\$L=30$ rounds}}
& \multirow{6}{*}{Tesseract}
& \multirow{3}{*}{$Z$}
& Monolithic & 23.0 & 2.3 & 2.5 & 2.8 & 3.2 & 3.9 & 5.3 & 8.4 & 19.3 & 48.5 & 279.0 & 17558.3 & $5.0\times10^{-6}$ & reference \\
\cline{4-18}
& & & \myCompilerName & 3.7 & 0.07 & 0.09 & 0.1 & 0.1 & 0.1 & 0.2 & 0.3 & 0.6 & 1.2 & 22.5 & 13582.7 & $5.8\times10^{-6}$ & 25/29 \\
\cline{4-18}
& & & Speedup & 6$\times$ & 32$\times$ & 29$\times$ & 28$\times$ & 26$\times$ & 27$\times$ & 28$\times$ & 30$\times$ & 33$\times$ & 42$\times$ & 12$\times$ & 1$\times$ & -- & -- \\
\cline{3-18}
& & \multirow{3}{*}{$X$}
& Monolithic & 21.0 & 2.3 & 2.5 & 2.7 & 3.0 & 3.6 & 4.8 & 7.6 & 17.0 & 42.0 & 255.3 & 15733.2 & $5.0\times10^{-6}$ & reference \\
\cline{4-18}
& & & \myCompilerName & 3.0 & 0.08 & 0.09 & 0.1 & 0.1 & 0.2 & 0.2 & 0.3 & 0.6 & 1.2 & 17.7 & 15642.5 & $3.2\times10^{-6}$ & 25/14 \\
\cline{4-18}
& & & Speedup & 7$\times$ & 29$\times$ & 27$\times$ & 25$\times$ & 23$\times$ & 24$\times$ & 24$\times$ & 26$\times$ & 28$\times$ & 36$\times$ & 14$\times$ & 1$\times$ & -- & -- \\
\hline
\multirow{6}{*}{\makecell[l]{HGP code\\$\llbracket 420,10,10\rrbracket$\\$L=30$ rounds}}
& \multirow{6}{*}{relay-BP}
& \multirow{3}{*}{$Z$}
& Monolithic & 32.5 & 16.4 & 18.0 & 20.5 & 23.0 & 25.7 & 30.5 & 39.2 & 60.7 & 85.0 & 175.8 & 1592.1 & $1.7\times10^{-7}$ & reference \\
\cline{4-18}
& & & \myCompilerName & 2.9 & 0.2 & 0.3 & 0.3 & 0.4 & 0.4 & 0.6 & 0.9 & 2.9 & 10.2 & 49.8 & 1592.1 & $0\,(<5.0\times10^{-7})$ & 28.08\%/12.99\% \\
\cline{4-18}
& & & Speedup & 11$\times$ & 75$\times$ & 70$\times$ & 68$\times$ & 64$\times$ & 58$\times$ & 53$\times$ & 43$\times$ & 21$\times$ & 8$\times$ & 4$\times$ & 1$\times$ & -- & -- \\
\cline{3-18}
& & \multirow{3}{*}{$X$}
& Monolithic & 32.6 & 15.7 & 18.1 & 20.7 & 23.2 & 25.9 & 30.8 & 38.6 & 61.2 & 85.6 & 176.2 & 1610.9 & $5.0\times10^{-7}$ & reference \\
\cline{4-18}
& & & \myCompilerName & 2.9 & 0.2 & 0.3 & 0.3 & 0.4 & 0.5 & 0.6 & 1.0 & 3.1 & 10.7 & 50.4 & 1610.9 & $1.7\times10^{-7}$ & 28.00\%/12.64\% \\
\cline{4-18}
& & & Speedup & 11$\times$ & 67$\times$ & 66$\times$ & 64$\times$ & 60$\times$ & 55$\times$ & 50$\times$ & 39$\times$ & 20$\times$ & 8$\times$ & 4$\times$ & 1$\times$ & -- & -- \\
\hline
\end{tabular}}
\end{table*}

For each configuration, we compare FusionQ with a monolithic invocation of the same backend on the same shot, classifying each shot's \myCompilerName\ cost as strictly \emph{lower} (L), strictly \emph{higher} (H), or tied against the monolithic cost.
For MWPM and Tesseract, L and H test whether localization preserves the minimum-weight objective; because Tesseract uses a finite search cap, a nonzero L or H can occur even though the backend is in principle exact, so we report both counts rather than presuming a perfect tie.
For heuristic relay-BP, which has no exact-minimum contract, the same L/H comparison instead reports how often localization returns a strictly cheaper or strictly more expensive correction than a monolithic relay-BP invocation.

Table~\ref{tab:overall-latency} summarizes the resulting latency distribution.
For each benchmark, we report the mean latency, latency percentiles from p20
through p99, and the worst observed latency.  We also append the logical error
rate and the L/H cost-consistency counts (exact backends) or percentages (relay-BP).
We report the fusion latency as $\min(T_{\mathrm{FusionQ}},T_{\mathrm{mono}})$ on each shot, corresponding to a race that publishes the first valid correction.
The monolithic path completes first for an average of $2.28\%$ of shots across the ten configurations, with a maximum of $4.41\%$ in any single configuration. Section~\ref{subsec:fusion-threshold} disables this backstop to expose FusionQ's raw crossover behavior.

\begin{figure*}[t]
\centering
\includegraphics[width=0.9\linewidth]{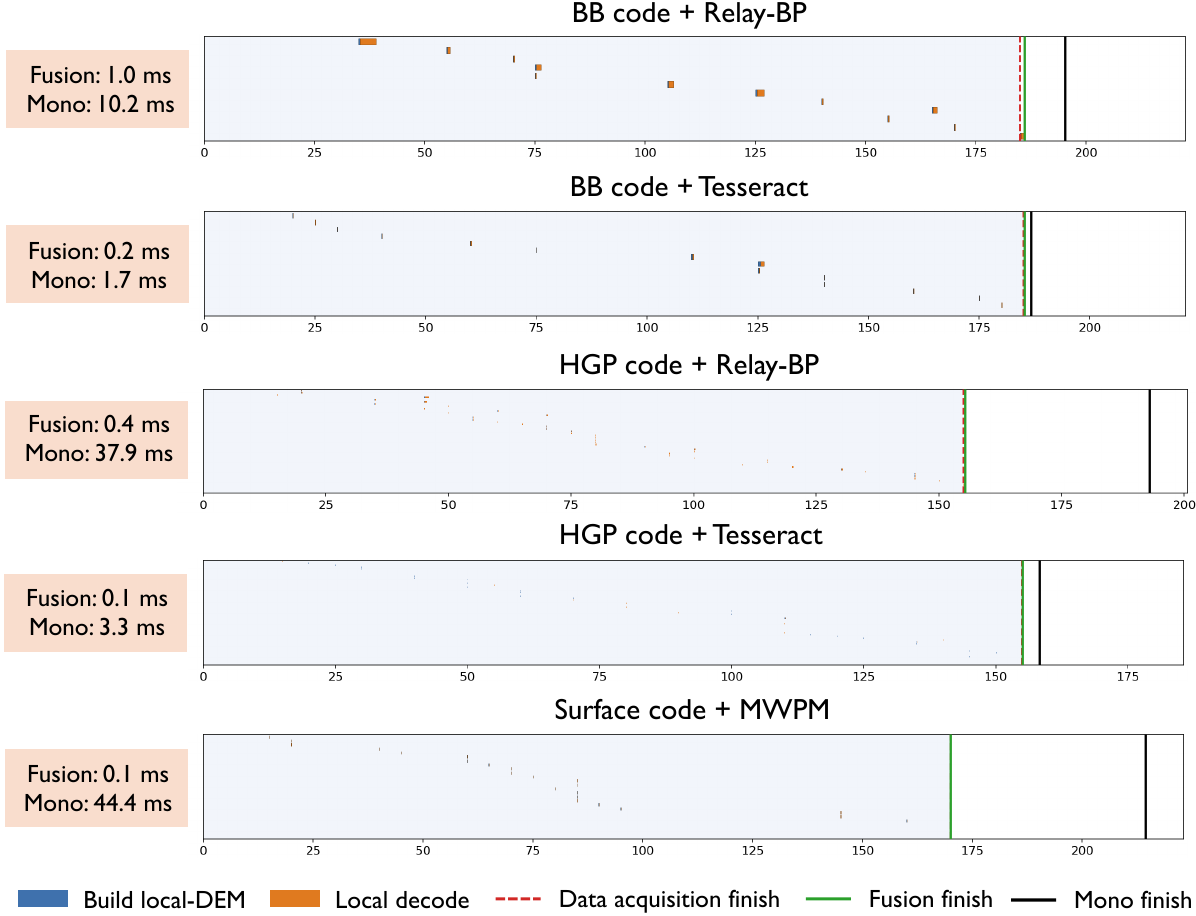}
\caption{Representative p50-shot latency timelines for the five overall-result
configurations ($Z$-basis shown). Each panel shows
syndrome acquisition, local DEM construction, local decoding, the acquisition frontier,
monolithic completion, and \myCompilerNameSpace publication.}
\label{fig:p90-latency-timelines}
\end{figure*}

Table~\ref{tab:overall-latency} shows that FusionQ keeps post-acquisition decoding below the $5$ms syndrome-production interval for a large majority of shots in every evaluated configuration.
For surface code with the MWPM decoder, the p99 latency is only $0.54$ms in both bases, so at least $99\%$ of corrections are published within $5$ms of the final syndrome arrival; the corresponding monolithic p99 latencies are $90.7$ and $90.4$ms.
For BB code with Tesseract decoder, the highest reported percentile below $5$ms is p99 in the $Z$ basis ($4.998$ms) and p95 in the $X$ basis ($1.46$ms); for HGP code with Tesseract decoder, p95 remains below the production interval in both bases ($1.16$ and $1.18$ms for Z and X bases respectively).
Across the exact-backend configurations, at least $95\%$ of corrections are published within $5$ms of the final syndrome arrival, and this fraction reaches at least $99\%$ in several configurations.
The extremely small fraction of cost discrepancies arises from capped Tesseract searches.

The relay-BP experiments demonstrate that the same fusion runtime can be used
with heuristic backends.  Since relay-BP does not provide an exact
minimum-weight contract, we do not expect it to tie the monolithic cost; instead, we report the L/H split directly.
With relay-BP, the highest reported percentile below $5$ms is p80 for the BB code and p90 for the HGP code, in both bases. The lower below-$5$ms percentiles reflect the heavier tail of the heuristic backend, while the mean post-acquisition decoding latency remains below $5$ms in all four relay-BP configurations. Across the two bases, FusionQ's correction is strictly cheaper than monolithic relay-BP on
$45.8\%$ of BB code shots and strictly more expensive on $16.9\%$; for HGP code the corresponding
figures are $28.0\%$ lower and $12.8\%$ higher (the remainder tie).

\begin{remark}
The absolute latency depends on the implementation efficiency of both the decoding backend and the FusionQ runtime. Because our evaluation uses a research prototype, a hardware-optimized implementation with specialized parallel processing and lower runtime overhead could further reduce the post-acquisition decoding latency.
\end{remark}

Figure~\ref{fig:p90-latency-timelines} explains the source of these latency
reductions. Monolithic decoding begins only after the full $L$-round syndrome
volume is available, whereas \myCompilerNameSpace overlaps local DEM
construction, footprint growth and union, and local decoding with syndrome
acquisition. In each timeline, the horizontal axis represents time, while
the vertical axis schematically distinguishes the regions affected by detected
errors and does not encode their absolute spatial locations. The blue bars show
the time spent constructing local detector error models, and the orange bars
show the corresponding local-decoding time. As a result, any local decoding
task that finishes before the final acquisition frontier contributes no
post-acquisition latency, even if it required boundary-dependent growth and
re-decoding. The remaining post-acquisition critical path consists only of the
live local decoding instances that have not finished by the time the last
syndrome data arrive. Consequently, post-acquisition latency is determined
by the remaining execution time of these unfinished tasks, rather than by the
total decoding work performed throughout syndrome acquisition.

\subsection{Scalability}
\label{subsec:scalability}

We next test whether FusionQ preserves its real-time behavior as the spatial and temporal decoding workloads grow. We show the experimental configurations in Table~\ref{tab:scalability-configs}.
The BB code-surgery sweep uses $p=2\times10^{-4}$ in the $X$ basis, while the surface-code sweeps use $p=5\times10^{-4}$ in the $Z$ basis, and every point contains $2$K shots with a $5$ms syndrome-production rate.
Because $2$K
shots are insufficient for reliable logical-error-rate estimates, 
we report the same L/H cost comparison as Section~\ref{subsec:overall-results}: for every
shot, whether \myCompilerNameSpace's cost is strictly lower (L) or strictly higher (H) than
monolithic's -- as raw shot counts for the exact backends (MWPM, Tesseract), and as percentages for
relay-BP.
In the code-surgery sweep, each data point uses GeneCS~\cite{zhou2026genecs} to generate a joint X-logical
measurement over multiple BB-code blocks, the number of physical qubits grows
linearly with the number of participating blocks.

\begin{table*}[t]
\centering
\caption{Experimental configurations for the scalability study.}
\label{tab:scalability-configs}
\small
\setlength{\tabcolsep}{5pt}
\renewcommand{\arraystretch}{1.15}
\resizebox{\textwidth}{!}{%
\begin{tabular}{lllcccc}
\toprule
\textbf{Study} &
\textbf{Workload} &
\textbf{Backend} &
\textbf{Swept parameter} &
\textbf{Physical error} &
\textbf{Shots} &
\textbf{Cost metric} \\
\midrule
Code surgery &
\makecell[l]{BB code\\$\llbracket 144,12,12\rrbracket$} &
Tesseract &
\makecell[c]{Blocks: $\{2,3,4,5,6\}$\\$L=36$} &
$2\times10^{-4}$ & $2$K &
L/H (count) \\
\midrule
Code surgery &
\makecell[l]{BB code\\$\llbracket 144,12,12\rrbracket$} &
relay-BP &
\makecell[c]{Blocks: $\{2,3,4,5,6\}$\\$L=36$} &
$2\times10^{-4}$ & $2$K &
L/H (\%) \\
\midrule
Distance scaling &
\makecell[l]{Surface code\\$\llbracket d^2,1,d\rrbracket$} &
MWPM &
\makecell[c]{Distance d: $\{11,13,15,21,31,41,51\}$\\$L=d$} &
$5\times10^{-4}$ & $2$K &
L/H (count) \\
\midrule
Round scaling &
\makecell[l]{Surface code\\$\llbracket 121,1,11\rrbracket$} &
MWPM &
\makecell[c]{Rounds L: $\{33,55,77,99,121,143,165\}$\\$d=11$} &
$5\times10^{-4}$ & $2$K &
L/H (count) \\
\bottomrule
\end{tabular}}
\end{table*}

For the surface-code sweeps, the monolithic baseline becomes impractical on the
largest instances.  We therefore run the monolithic baseline only through
$d=31$ in the distance-scaling study and
for every point in the round-scaling study;
the $d=41$ and $d=51$ distance-scaling points are therefore reported for FusionQ only.

\begin{figure*}[t]
\centering
\includegraphics[width=0.9\linewidth]{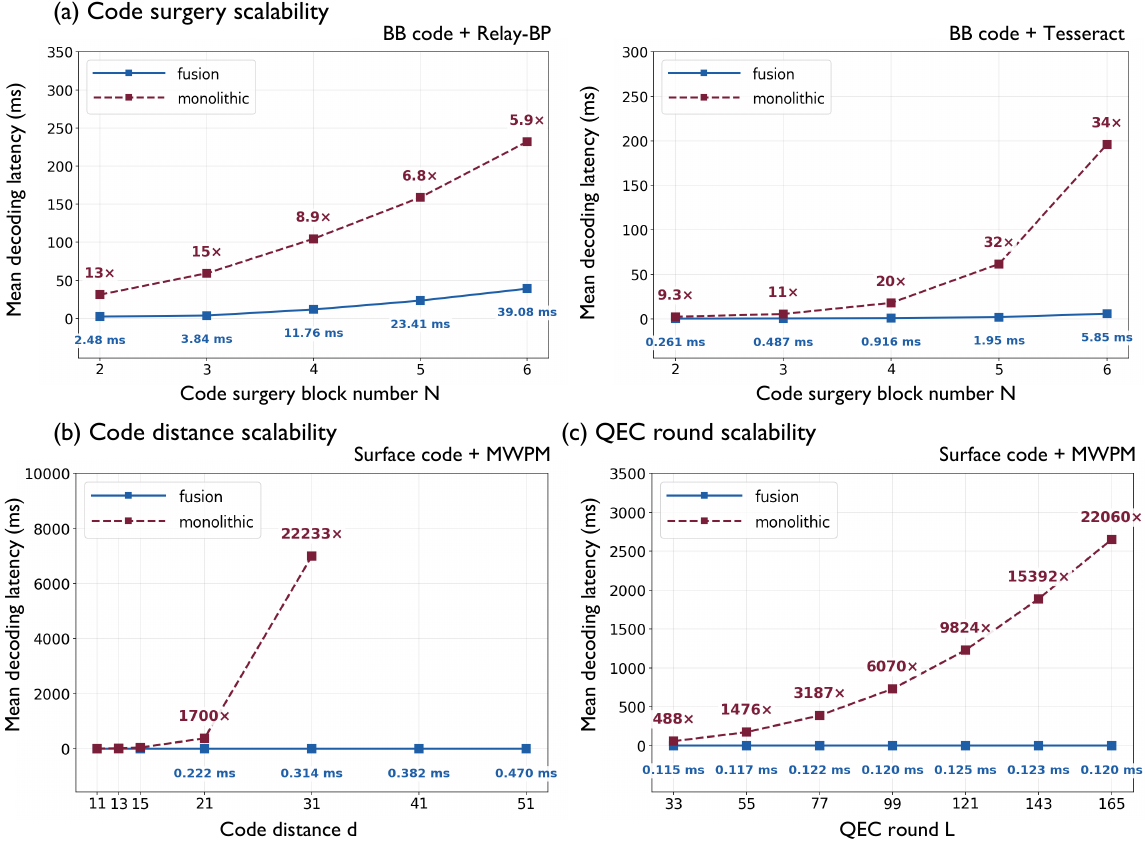}
\caption{Scalability of mean post-acquisition latency.}
\label{fig:scalability}
\end{figure*}

Figure~\ref{fig:scalability} shows that FusionQ's residual critical path grows much more slowly than the full decoding instance.
In the BB-code surgery experiment with Tesseract decoder, increasing the number of
merged blocks from $2$ to $6$ causes the monolithic latency to grow from
2.43 ms to 196.04 ms, while \myCompilerNameSpace stays
between 0.26 ms and 5.85 ms.  This shows that
\myCompilerNameSpace does not pay the full cost of decoding the merged
code-surgery DEM as one global instance.
Even at six blocks, FusionQ's $5.85$ms mean is approximately one syndrome-production interval, whereas monolithic decoding adds $196$ms after acquisition.
Across the five block counts, FusionQ's cost ties the monolithic cost on every shot through
four blocks, and sees only $1$ (five blocks) and $3$ (six blocks) strictly-lower-cost shots out of
$2$K, with no strictly-higher-cost shot at any block count (L/H of $0/0$, $0/0$, $0/0$, $1/0$,
$3/0$). With relay-BP at six blocks, FusionQ reduces mean latency from $232.08$ to $39.08$ms
($5.94\times$); its correction is strictly cheaper on $59.3\%$ of shots and strictly more expensive
on $33.65\%$.

The same trend appears in the surface-code distance sweep.  The monolithic
baseline grows rapidly with distance and becomes impractical beyond $d=31$,
whereas \myCompilerNameSpace continues through $d=51$ with mean latency
increasing only from 0.09 ms to 0.47 ms.  Across the
points where the monolithic exact baseline is available ($d\le31$), \myCompilerNameSpace ties the monolithic cost on every shot (L/H of $0/0$ at every distance).

Finally, the round-scaling experiment isolates the effect of temporal depth at
fixed distance $d=11$.
Monolithic mean latency grows from $56.2$ms at $L=33$ to $2650.7$ms at $L=165$ because all rounds enter one end-of-acquisition solve.
FusionQ keeps mean post-acquisition latency within $0.115$--$0.125$ms over the same $5\times$ increase in circuit depth, tying the monolithic cost on every shot (L/H of $0/0$) at every point and a $22{,}060\times$ speedup at $L=165$.
This depth-independent residual is the strongest evidence that FusionQ operates as a streaming real-time decoder: additional rounds provide additional acquisition time in which their local decoding work is completed, rather than accumulating an end-of-circuit backlog.

\subsection{Comparison with Parallel Window Decoding}
\label{subsec:window-comparison}

We next compare FusionQ with parallel window decoding to isolate the effect of scheduling granularity between two methods that both overlap decoding with syndrome acquisition.
We use memory experiments so that the comparison does not depend on a particular logical-operation layout.
All experiments use
physical error rate $p=5\times10^{-4}$, $2$K shots, and $5$ms
syndrome-acquisition time per QEC round.  Since $2$K shots are not enough to
estimate stable logical error rates, we instead report latency percentiles, each decoder's logical-error \emph{count} against ground-truth observable flip.

\begin{table}[t]
\centering
\caption{Experimental configurations for the comparison with parallel window decoding.}
\label{tab:window-configs}
\small
\setlength{\tabcolsep}{5pt}
\renewcommand{\arraystretch}{1.08}
\resizebox{\linewidth}{!}{%
\begin{tabular}{llcccccc}
\toprule
\textbf{Benchmark} &
\textbf{Code} &
\textbf{Rounds} &
\textbf{Backend} &
\textbf{Physical error} &
\textbf{Shots} &
\textbf{Acq./round} &
\textbf{Cost metric} \\
\midrule
BB code & $\llbracket 144,12,12\rrbracket$ &$L=kd,\ k\in\{19,23,27,31,35\}$ & Tesseract & $5\times10^{-4}$ & $2$K & $5$ms & Error count \\
Surface code & $\llbracket 121,1,11\rrbracket$ & $L=kd,\ k\in\{19,23,27,31,35\}$ & MWPM & $5\times10^{-4}$ & $2$K & $5$ms & Error count \\
\bottomrule
\end{tabular}}
\end{table}

\begin{table*}[t]
\centering
\caption{Mean and p95 post-acquisition latency (ms) and logical-error counts across the $L=kd$ sweep.}
\label{tab:window-latency}
\scriptsize
\setlength{\tabcolsep}{4pt}
\renewcommand{\arraystretch}{1.1}
\resizebox{\textwidth}{!}{%
\begin{tabular}{|c|c|c|c|c|c|c|c|c|c|}
\hline
\multirow{2}{*}{\textbf{Benchmark}} & \multirow{2}{*}{\textbf{Basis}} & \multirow{2}{*}{\textbf{$k$}} &
\multicolumn{3}{c|}{\textbf{mean (ms)}} & \multicolumn{3}{c|}{\textbf{p95 (ms)}} &
\multirow{2}{*}{\textbf{\#fus / \#win err}} \\
\cline{4-9}
& & & \textbf{window} & \textbf{\myCompilerName} & \textbf{speedup} & \textbf{window} & \textbf{\myCompilerName} & \textbf{speedup} & \\
\hline
\multirow{10}{*}{\makecell[l]{BB code\\$\llbracket 144,12,12\rrbracket$\\Tesseract}}
& \multirow{5}{*}{$Z$} & 19 & 31.47 & 15.23 & 2$\times$ & 21.39 & 1.59 & 13$\times$ & 0/1 \\
\cline{3-10}
& & 23 & 29.85 & 13.86 & 2$\times$ & 18.89 & 1.47 & 13$\times$ & 0/1 \\
\cline{3-10}
& & 27 & 41.08 & 19.20 & 2$\times$ & 22.01 & 1.65 & 13$\times$ & 0/1 \\
\cline{3-10}
& & 31 & 48.50 & 23.67 & 2$\times$ & 28.42 & 1.65 & 17$\times$ & 0/2 \\
\cline{3-10}
& & 35 & 56.84 & 24.71 & 2$\times$ & 25.83 & 2.03 & 13$\times$ & 0/3 \\
\cline{2-10}
& \multirow{5}{*}{$X$} & 19 & 139.62 & 3.78 & 37$\times$ & 37.82 & 2.17 & 17$\times$ & 0/8 \\
\cline{3-10}
& & 23 & 152.74 & 3.32 & 46$\times$ & 56.04 & 1.79 & 31$\times$ & 0/9 \\
\cline{3-10}
& & 27 & 179.44 & 3.35 & 53$\times$ & 59.65 & 1.91 & 31$\times$ & 0/11 \\
\cline{3-10}
& & 31 & 186.63 & 3.09 & 60$\times$ & 77.52 & 2.21 & 35$\times$ & 0/12 \\
\cline{3-10}
& & 35 & 234.49 & 4.25 & 55$\times$ & 132.38 & 2.42 & 55$\times$ & 0/16 \\
\hline
\multirow{10}{*}{\makecell[l]{Surface code\\$\llbracket 121,1,11\rrbracket$\\MWPM}}
& \multirow{5}{*}{$Z$} & 19 & 124.06 & 0.12 & 997$\times$ & 161.99 & 0.25 & 651$\times$ & 0/0 \\
\cline{3-10}
& & 23 & 123.17 & 0.12 & 1009$\times$ & 160.24 & 0.25 & 646$\times$ & 0/0 \\
\cline{3-10}
& & 27 & 125.21 & 0.15 & 862$\times$ & 162.23 & 0.27 & 600$\times$ & 0/0 \\
\cline{3-10}
& & 31 & 124.06 & 0.13 & 970$\times$ & 161.92 & 0.27 & 591$\times$ & 0/0 \\
\cline{3-10}
& & 35 & 126.21 & 0.13 & 943$\times$ & 164.38 & 0.28 & 577$\times$ & 0/0 \\
\cline{2-10}
& \multirow{5}{*}{$X$} & 19 & 124.08 & 0.12 &1010$\times$ & 161.30 & 0.25 & 652$\times$ & 0/0 \\
\cline{3-10}
& & 23 & 123.26 & 0.12 & 993$\times$ & 159.71 & 0.25 & 650$\times$ & 0/0 \\
\cline{3-10}
& & 27 & 125.33 & 0.14 & 866$\times$ & 162.78 & 0.26 & 617$\times$ & 0/0 \\
\cline{3-10}
& & 31 & 130.93 & 0.13 & 993$\times$ & 191.13 & 0.29 & 657$\times$ & 0/0 \\
\cline{3-10}
& & 35 & 126.21 & 0.13 & 939$\times$ & 164.11 & 0.28 & 583$\times$ & 0/0 \\
\hline
\end{tabular}}
\end{table*}

The QEC depth is set to $L=kd$ for $k\in\{19,23,27,31,35\}$
in both benchmarks. Under the parallel
window-decoding schedule, $L=kd$ corresponds to $\lceil (k+1)/4\rceil$ $A$ windows and $\lceil (k+1)/4\rceil - 1$ $B$
windows, where each window spans $3d$ rounds; $k=19,\ldots,35$ therefore gives
$5,6,7,8,9$ $A$ windows and $4,5,6,7,8$ $B$ windows, respectively. Both parallel window decoding and \myCompilerNameSpace
use the same decoding backend.

Table~\ref{tab:window-latency} shows the results. FusionQ leaves a substantially shorter post-acquisition critical path than parallel window decoding throughout the $k=19,\ldots,35$ sweep. Parallel window decoding can hide completed window solves behind syndrome acquisition, but each window still spans a fixed $3d$ rounds. FusionQ instead forms smaller syndrome-induced local decoding tasks, leaving much less work on the critical path after acquisition.

\begin{figure*}[t]
\centering
\includegraphics[width=0.9\textwidth]{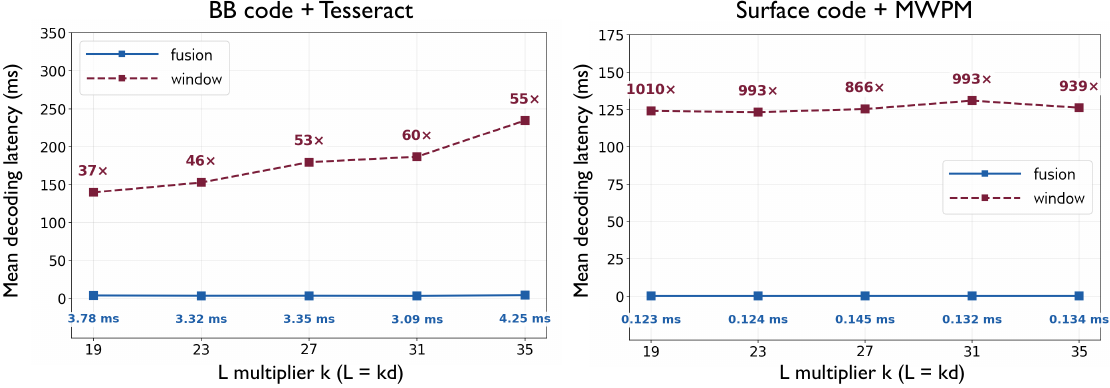}
\caption{Mean post-acquisition latency vs.\ QEC depth $L=kd$ (X-basis shown).}
\label{fig:window-sweep}
\end{figure*}

\begin{figure*}[t]
\centering
\includegraphics[width=0.9\textwidth]{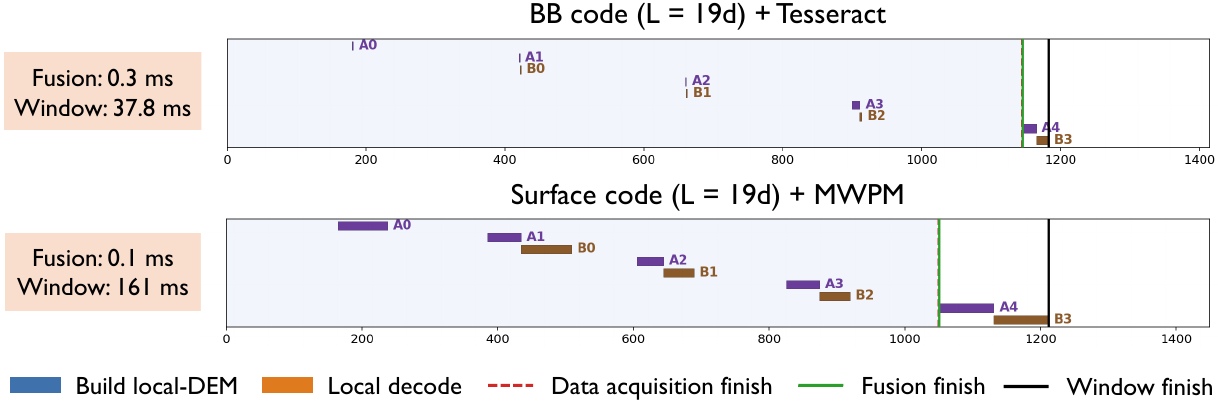}
\caption{Representative p95-shot latency timelines of window decoding (X-basis shown).}
\label{fig:window-p90-latency}
\end{figure*}

FusionQ's p95 post-acquisition latency stays below the $5$ms syndrome-production interval at every tested depth: $1.47$--$2.42$ms for BB code with Tesseract and $0.25$--$0.29$ms for surface code with MWPM. At $k=19$ in the $Z$ basis, FusionQ reduces p95 latency from $21.39$ms to $1.59$ms ($13.4\times$) for the BB code, and from $161.99$ms to $0.25$ms ($650.6\times$) for the surface code. The BB/Tesseract mean can exceed its p95 because a small number of unusually slow shots disproportionately raise the average; the p95 results still show that at least $95\%$ of shots finish within one syndrome production interval. Across the sweep, BB-code mean speedup ranges from $2.0$--$2.3\times$ in the $Z$ basis to $36.9$--$60.4\times$ in the $X$ basis, while surface-code mean speedup remains above $860\times$.

Neither decoder records a logical error in the surface-code experiments. For the BB code, FusionQ records no logical errors in either basis at any tested depth, whereas parallel window decoding records 1--3 errors in the $Z$ basis and 8--16 in the $X$ basis. Some of the observed errors may be attributable to Tesseract's capped search with \texttt{pqlimit}$=10^7$.


Figure~\ref{fig:window-p90-latency} visualizes the p95-shot latency timelines for the BB-code and surface-code benchmarks at $k=19$. Each panel shows the individual $3d$-round window solves, indicating which finish during syndrome acquisition and which remain on the post-acquisition critical path. The figure also marks FusionQ's completion time for the same shot, without displaying its individual local decoding tasks.


\subsection{Fusion Decoding Threshold}
\label{subsec:fusion-threshold}

Finally, we study how FusionQ's latency acceleration scales with code size as the physical error rate changes.
As the physical error rate rises, fired syndrome events become denser, causing local clusters to grow and merge more frequently.
We define the fusion-decoding threshold as the physical error rate at which the latency-acceleration curves for different code sizes intersect.
Below this threshold, larger code instances achieve greater acceleration over monolithic decoding; above it, this favorable size dependence disappears or reverses.
This crossing behavior is analogous to the conventional code threshold obtained from logical-error curves across code distances, but characterizes decoding-latency scaling rather than logical reliability.
To identify the threshold for each code--backend pair, we sweep four BB-code sizes and four surface-code distances, setting the circuit depth to $L=3d$ for every configuration.

All experiments use $5$ms syndrome-acquisition time per QEC round.
Unlike the preceding studies, this sweep reports raw FusionQ latency without the monolithic race, because applying the per-shot minimum would clip the size-dependent acceleration curves and obscure their intersection.
We report mean latency to expose the aggregate transition from sparse, independently solvable clusters to large merged decoding tasks.

\begin{table}[t]
\centering
\caption{Experimental configurations for the fusion decoding threshold study.}
\label{tab:threshold-configs}
\small
\setlength{\tabcolsep}{4pt}
\renewcommand{\arraystretch}{1.08}
\resizebox{\linewidth}{!}{%
\begin{tabular}{llccccc}
\toprule
\textbf{Benchmark} &
\textbf{Code} &
\textbf{Rounds} &
\textbf{Backend} &
\textbf{Physical error} &
\textbf{Acq./round} &
\textbf{Shots} \\
\midrule
BB code & $\llbracket 72,12,6\rrbracket$ & $L=18$ & Tesseract & $\{2,3,4,5,6,7,8\}\times10^{-4}$ & $5$ms & $10$K \\
BB code & $\llbracket 90,8,10\rrbracket$ & $L=30$ & Tesseract & $\{2,3,4,5,6,7,8\}\times10^{-4}$ & $5$ms & $10$K \\
BB code & $\llbracket 108,8,10\rrbracket$ & $L=30$ & Tesseract & $\{2,3,4,5,6,7,8\}\times10^{-4}$ & $5$ms & $10$K \\
BB code & $\llbracket 144,12,12\rrbracket$ & $L=36$ & Tesseract & $\{2,3,4,5,6,7,8\}\times10^{-4}$ & $5$ms & $10$K ($1$K for $p\geq7\times10^{-4}$) \\
\midrule
BB code & $\llbracket 72,12,6\rrbracket$ & $L=18$ & relay-BP & $\{2,4,6,8,10\}\times10^{-4}$ & $5$ms & $10$K \\
BB code & $\llbracket 90,8,10\rrbracket$ & $L=30$ & relay-BP & $\{2,4,6,8,10\}\times10^{-4}$ & $5$ms & $10$K \\
BB code & $\llbracket 108,8,10\rrbracket$ & $L=30$ & relay-BP & $\{2,4,6,8,10\}\times10^{-4}$ & $5$ms & $10$K \\
BB code & $\llbracket 144,12,12\rrbracket$ & $L=36$ & relay-BP & $\{2,4,6,8,10\}\times10^{-4}$ & $5$ms & $10$K \\
\midrule
Surface code & $\llbracket 25,1,5\rrbracket$ & $L=15$ & MWPM & $\{1,2,3,4,5,6\}\times10^{-3}$ & $5$ms & $1$K \\
Surface code & $\llbracket 49,1,7\rrbracket$ & $L=21$ & MWPM & $\{1,2,3,4,5,6\}\times10^{-3}$ & $5$ms & $1$K \\
Surface code & $\llbracket 81,1,9\rrbracket$ & $L=27$ & MWPM & $\{1,2,3,4,5,6\}\times10^{-3}$ & $5$ms & $1$K \\
Surface code & $\llbracket 121,1,11\rrbracket$ & $L=33$ & MWPM & $\{1,2,3,4,5,6\}\times10^{-3}$ & $5$ms & $1$K ($100$ for $p\geq4\times10^{-3}$) \\
\bottomrule
\end{tabular}}
\end{table}

\begin{figure*}[t]
\centering
\includegraphics[width=1.0\textwidth]{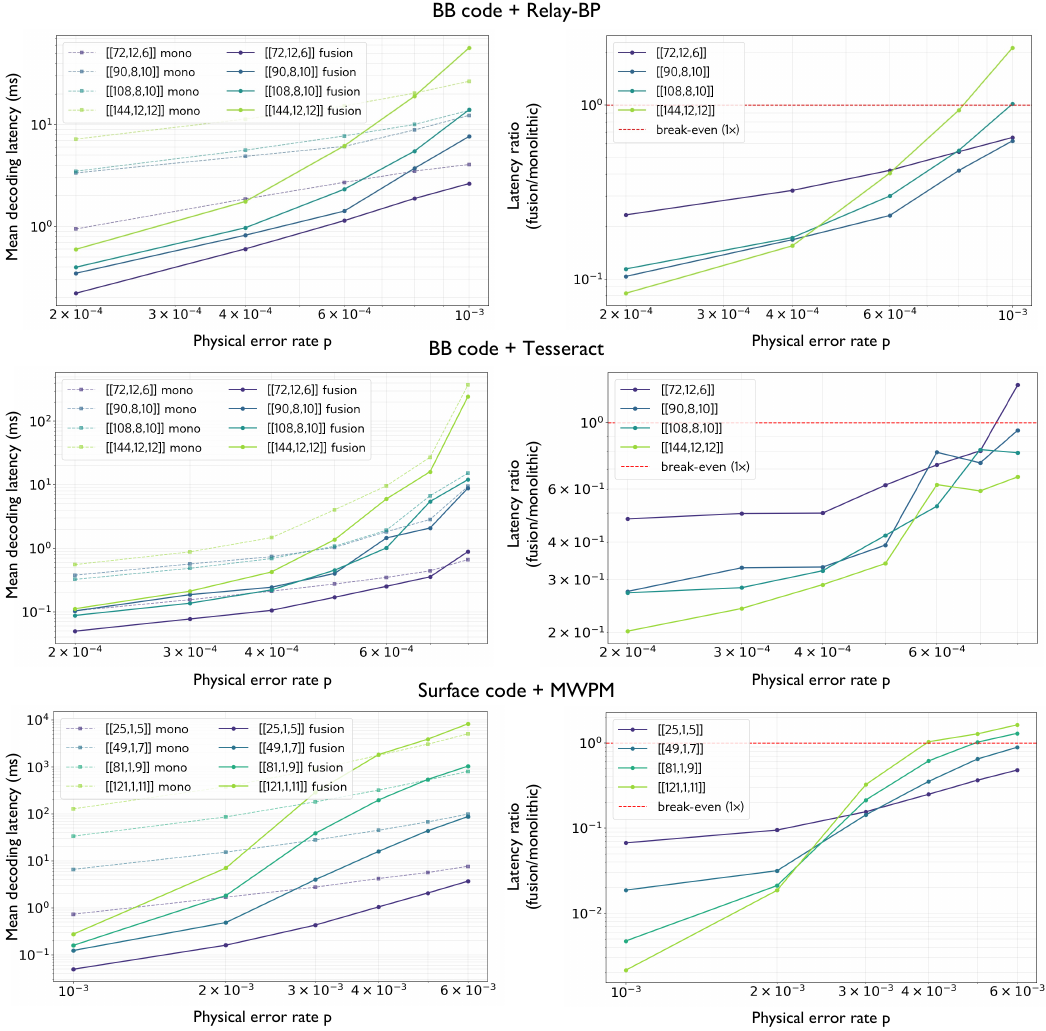}
\caption{Mean post-acquisition latency (left) and acceleration over monolithic decoding (right) versus physical error rate ($Z$ basis). Rows show BB/relay-BP and BB/Tesseract ($n\in\{72,90,108,144\}$), followed by surface/MWPM ($d\in\{5,7,9,11\}$). Crossings among the size-dependent acceleration curves provide empirical threshold estimates; the red dashed $1\times$ line separately marks latency break-even.}
\label{fig:fusion-threshold}
\end{figure*}

Figure~\ref{fig:fusion-threshold} shows the raw post-acquisition latency and the corresponding acceleration over monolithic decoding.
At low physical error rates, sparse syndrome events allow FusionQ to exploit locality increasingly effectively as the code grows, so larger instances achieve greater acceleration.
As the error rate rises, local tasks grow and merge more frequently, and the acceleration curves for different code sizes converge and cross.
These common crossing regions occur near $p\approx5\times10^{-4}$ for BB/relay-BP, $p\approx7\times10^{-4}$ for BB/Tesseract, and $p\approx3\times10^{-3}$ for surface/MWPM.
Below each crossing, increasing code size improves FusionQ's acceleration; above it, the size ordering disappears or reverses.
The crossings among the size-dependent curves provide empirical threshold estimates, whereas the red dashed $1\times$ line indicates latency break-even with monolithic decoding. These crossings are finite-size empirical observations; whether they converge to a well-defined threshold as code size grows, and whether local cluster sizes have a uniform exponential tail below it, remain open questions (Section~\ref{sec:future}).


\section{Related Work}
\label{sec:related}

\paragraph{Window-based decoding.}
Parallel-window~\cite{skoric2023parallel}, sandwich~\cite{tan2023scalable},
spatial-window~\cite{lin2025spatially}, and modular
decoding~\cite{bombin2023modular} support interchangeable local solvers with
syndrome-agnostic window configurations.
SWIPER~\cite{viszlai2025swiper} speculates on boundary information to start
dependent windows earlier, while ADaPT~\cite{oberoi2026adapt} and
complementary-gap methods~\cite{mishima2026adaptive} adapt window or buffer
sizes to decoding confidence.

\paragraph{Fusion and decoder interfaces.}
Fusion Blossom combines matching-specific primal-dual states to recover global
MWPM~\cite{wu2023fusion}. Micro Blossom demonstrates low-latency hardware
fusion~\cite{wu2025micro}, while Google's experiment uses
finite-history fusion~\cite{google2025quantum}. Fusion has also been developed
for Union-Find in DecoNet~\cite{liyanage2025network} and explored for
Minimum-Weight Parity Factor (MWPF) decoder~\cite{yang2024parallel}. LEGO proposes
fusion-capable decoding blocks but leaves adaptation of further backends
open~\cite{wu2024lego}. \myCompilerNameSpace separates fusion from these
solver-specific mechanisms: its local-DEM interface requires only a selected
error set and a virtual-boundary-use flag. Our implementation retains cluster
footprints, ownership, and tentative corrections but performs fresh local
solves; incremental backend-state evolution is a possible optimization, not an
implemented feature or a correctness requirement.

\paragraph{Syndrome-dependent clustering.}
Union-Find variants~\cite{delfosse2021almost,delfosse2022toward,cain2024correlated},
localized statistics decoding (LSD)~\cite{hillmann2025localized}, and Ambiguity
Clustering~\cite{wolanski2024ambiguity} exploit syndrome-dependent regions.
Local solvability alone, however, does not certify global minimum weight.
Continuous cluster processing also predates \myCompilerName: Snowflake retains
state across streaming windows~\cite{chan2026snowflake}, while Cluster-As-You-Go
grows clusters during acquisition and commits neutral clusters after a
configurable lifetime~\cite{peduzzi2026reducing}. \myCompilerNameSpace keeps
boundary-free local answers tentative, allowing later absorption and
re-decoding. To our knowledge, \myCompilerNameSpace is the first solver-parametric
framework for general DEMs to combine syndrome-driven dynamic clustering with
a global minimum-weight guarantee when the local backend is exact.

\paragraph{Parallel-runtime guarantees.}
Several MWPM decoders exploit locality in
matching~\cite{fowler2013minimum,higgott2025sparse,wu2023fusion}. Recent work proves a finite
Union-Find threshold with quasi-polylogarithmic average parallel
runtime~\cite{yoshida2026proof}. Mikami and Yamasaki construct syndrome-derived
clusters that reproduce global Sparse Blossom in quasi-polylogarithmic average
total parallel time for a $d$-round window, yielding $o(1)$ average time per
round~\cite{mikami2026sparse}. These proofs rely on decoder-specific dynamics
and surface-code locality, but their cluster-based techniques offer a promising
starting point for analyzing \myCompilerName. Extending them to our dynamic
cluster framework remains future work. Our composition
theorem establishes global optimality, not a runtime bound.

\section{Future Directions}
\label{sec:future}

Our results motivate two directions for future work: a hardware implementation of FusionQ for faster QEC platforms, and a theoretical characterization of the fusion-decoding threshold through the statistics of local clusters.

\subsection{Hardware-Accelerated Implementation}

Our evaluation uses a research prototype: local decoding tasks are timed on one
CPU core and replayed under an ideal parallel-resource model. We consider a
$5\,\mathrm{ms}$ syndrome-production period, representative of millisecond-scale
QEC cycles reported for neutral-atom hardware and assumed for trapped-ion
architectures~\cite{bluvstein2026fault,loschnauer2025scalable,hughes2025trapped}.
An end-to-end implementation on FPGAs, GPUs, or ASICs could parallelize local-DEM
construction, scheduling, footprint growth and union, and backend decoding across
clusters and within each solve, thereby reducing both typical and tail
post-acquisition latency. Micro Blossom demonstrates that fusion can be supported by FPGA hardware in a CPU--FPGA decoder specialized for MWPM~\cite{wu2025micro}, providing a concrete precedent for hardware acceleration of FusionQ's solver-parametric fusion framework.

Such acceleration could extend FusionQ to faster hardware, including
superconducting processors with syndrome-production periods of approximately
$1\,\mu\mathrm{s}$~\cite{google2025quantum}. Meeting this target requires
accelerating the entire path from syndrome arrival to correction publication,
rather than only the backend-decoder kernel. Future hardware evaluations should
therefore report end-to-end latency distributions, resource utilization, and
communication costs, emphasizing the high-percentile and worst-case behavior
that determines whether decoding backlog accumulates.

\subsection{A Theory of the Fusion-Decoding Threshold}

Section~\ref{subsec:fusion-threshold} identifies the fusion-decoding threshold
empirically through crossings of size-dependent latency-acceleration curves. A
theoretical characterization should relate this behavior to the statistics of
the syndrome-induced clusters and determine when FusionQ has constant expected
local decoding cost.

Let $N$ denote the size of the global decoding instance, and let $C$ denote a
FusionQ cluster generated at physical error rate $p$, with $|C|$ equal to the
number of detectors in its footprint.

\begin{hypothesis}[Uniform exponential cluster-size tail]
\label{hyp:fusion-threshold}
For a fixed code family $\mathcal{Q}$ and decoding backend $\mathcal{B}$, there
exists a fusion-decoding threshold
$p_{\mathrm{fus}}(\mathcal{Q},\mathcal{B})$ such that, for every
$p<p_{\mathrm{fus}}(\mathcal{Q},\mathcal{B})$, there are finite constants
$A_{\mathcal{Q},\mathcal{B}}(p)$ and
$\xi_{\mathcal{Q},\mathcal{B}}(p)$ satisfying
\[
\begin{gathered}
\Pr_p\!\left[|C|\ge s\right]
\le
A_{\mathcal{Q},\mathcal{B}}(p)
e^{-s/\xi_{\mathcal{Q},\mathcal{B}}(p)},
\end{gathered}
\]
for all $s\ge0$. These constants may depend on the structural properties of the code family,
including its expansion, and on the decoding backend, but are independent of
the global instance size $N$.
\end{hypothesis}

Suppose that the worst-case runtime of the inner decoder on a cluster of size
$s$ grows at most polynomially,
\[
T_{\mathrm{in}}(s)\le c s^\alpha,
\]
with $c>0$ and $\alpha\ge0$. Hypothesis~\ref{hyp:fusion-threshold} would then imply a finite bound
\[
\mathbb{E}_p\!\left[T_{\mathrm{in}}(|C|)\right]
\le K(p,\alpha)<\infty,
\]
where $K(p,\alpha)$ is independent of the global instance size $N$.

Consequently, increasing the code size below the threshold would not increase
the expected cost of an individual local solve, whereas monolithic decoding
must process a global instance that grows with $N$. This contrast explains why
FusionQ's acceleration becomes more favorable for larger codes without implying
that its absolute latency decreases with code size. As $p$ approaches
$p_{\mathrm{fus}}$, the characteristic cluster size $\xi_{\mathcal{Q},\mathcal{B}}(p)$ may grow, weakening
this advantage and producing the size-dependent acceleration crossings observed
in Section~\ref{subsec:fusion-threshold}. Proving
Hypothesis~\ref{hyp:fusion-threshold} and characterizing
$p_{\mathrm{fus}}$ remain important directions for future work.


\bibliographystyle{ACM-Reference-Format}
\bibliography{main}



\end{document}